\documentclass[a4paper,twocolumn,11pt,accepted=2023-05-08]{quantumarticle}
\pdfoutput=1
\usepackage[utf8]{inputenc}
\usepackage[english]{babel}
\usepackage[T1]{fontenc}
\usepackage{amsmath}
\usepackage{amssymb}
\usepackage{hyperref}
\usepackage[numbers,sort&compress]{natbib}

\usepackage{tikz}
\usepackage{lipsum}

\begin{document}

\title{Driven-dissipative topological phases in parametric resonator arrays}
\author{\'{A}lvaro G\'{o}mez-Le\'{o}n}
\email{a.gomez.leon@csic.es}
\affiliation{Instituto de F\'isica Fundamental (IFF), CSIC, Calle Serrano 113b, 28006 Madrid, Spain.}

\author{Tom\'{a}s Ramos}
\email{t.ramos.delrio@gmail.com}
\affiliation{Instituto de F\'isica Fundamental (IFF), CSIC, Calle Serrano 113b, 28006 Madrid, Spain.}

\author{Alejandro Gonz\'{a}lez-Tudela}
\email{a.gonzalez.tudela@csic.es}
\affiliation{Instituto de F\'isica Fundamental (IFF), CSIC, Calle Serrano 113b, 28006 Madrid, Spain.}

\author{Diego Porras}
\email{diego.porras@csic.es}
\affiliation{Instituto de F\'isica Fundamental (IFF), CSIC, Calle Serrano 113b, 28006 Madrid, Spain.}

\maketitle

\begin{abstract}
We study the phenomena of topological amplification in arrays of parametric oscillators.
We find two phases of topological amplification, both with directional transport and exponential gain with the number of sites, and one of them featuring squeezing. 
We also find a topologically trivial phase with zero-energy modes which produces amplification but lacks the robust topological protection of the others.
We characterize the resilience to disorder of the different phases and their stability, gain, and noise-to-signal ratio. Finally, we discuss their experimental implementation with state-of-the-art techniques.
\end{abstract}

\maketitle
\section{Introduction}
Systems with topological properties have become a cornerstone in the development of current technologies. Since the discovery of the quantum Hall effect~\cite{QHE1,QHE2}, which has allowed to measure physical constants with high accuracy~\cite{VonKlitzing-metrology}, the synthesis of graphene~\cite{Novoselov2007}, celebrated with the Nobel prize in physics, and the successive discovery of spin topological insulators~\cite{Bernevig-STI}, we are now at a time where topologically protected systems are present in a vast number of fields. 

A promising area where topology can play a pivotal role is photonics.
There, it can be used to produce exotic couplings between quantum emitters~\cite{Tudela2019,Painter2021,Barik2018,Vega2021,Elcano2020,Elcano2021,Leonforte2021,DeBernadis2021}, to transport and manipulate light~\cite{Photonic-TI,Photonic-TI2,TopologicalPhotonics,Advances-Top-Phot,Photonic3D-realization,TopologicalPhotonics2,2DTopPhotonics,Perspective-TopPhotonics} or to improve sensing capabilities in metrology~\cite{Sensing1,Sensing-McDonald2020,Sensing-Koch2021,Sensing2021}. 
Importantly, topological phases in photonic systems typically include gain and loss, and this makes them fundamentally different to the ones typically considered in materials science~\cite{TI-Dissipation}.
In fact, the dissipative nature of the systems can enrich the physics of the topological phases~\cite{PhysRevX.8.031079,PhysRevX.9.041015,Periodic-table-Non-hermitian,Skin-Effect-Invariants,McDonaldPRB2022,PhysRevLett.124.056802}, and this can be detected in the steady state~\cite{Nunnenkamp2020,PhysRevA.103.033513} or in the transient properties~\cite{Viola3,AGL-Decimation,Skin-Effect-Damping}.
An interesting case of a dissipative topological phase, present in photonic systems, is that of topological amplification~\cite{Clerk2016,McDonald2018,PhysRevLett.122.143901,Nunnenkamp2020,PhysRevA.103.033513}.
There, the topological nature of the open quantum system induces perfectly directional amplification of signals with an exponential gain, near quantum limited noise and robustness to disorder~\cite{Wanjura2021}.
\begin{figure}
    \centering
    \includegraphics[width=1\columnwidth]{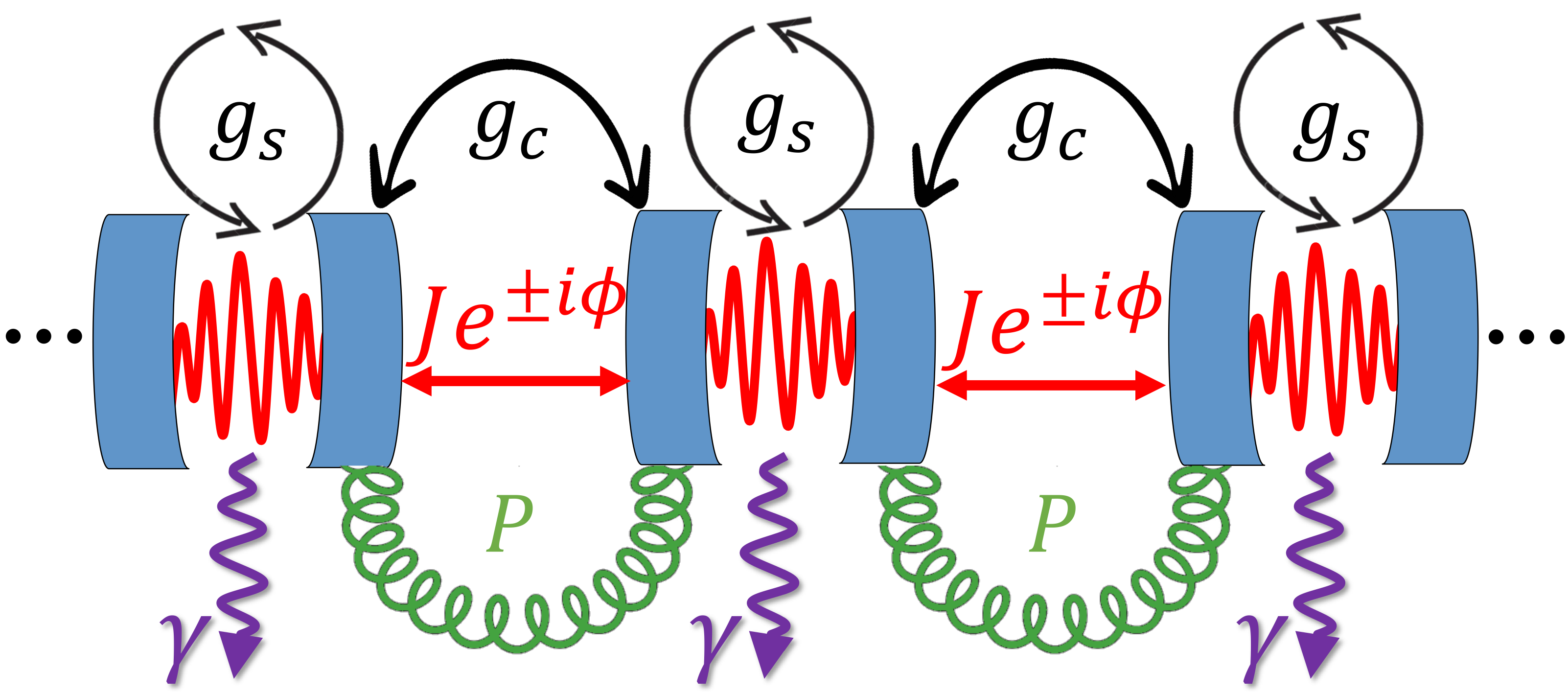}
    \caption{Schematic of a topological TWPA showing the coherent hopping in red, the parametric terms in black and the dissipative ones in purple and green. The coherent contributions include a hopping with phase $\phi$ and squeezing terms, $g_{s,c}$, arising from parametric driving of the cavities. The dissipative terms describe local dissipation $\gamma$ and collective pump $P$.}
    \label{fig:Schematic}
\end{figure}

In this work, we explore the amplification properties of an array of coupled parametric oscillators (see Fig.~\ref{fig:Schematic} for a schematic description of the system). 
This traveling-wave parametric amplifier~(TWPA)~\cite{Cullen1958,TWPA-josephson,TWPA-QuantumLimitted} features topologically protected amplification against all types of disorder. We analyze the experimental requirements to enter the topological phase, the amplification properties and its robustness to disorder.

Our proposed setup has some important differences with previously considered models~\cite{peano2016,McDonald2018}. Here we study a one-dimensional chain with both local and collective parametric driving terms, and include the presence of homogeneous dissipation, which acts all over the array.
This makes the bulk of the topological phase stable, in contrast to 2D proposals which consider local sinks to stabilize the propagating edge states~\cite{peano2016}, and also leads to strong robustness against all types of disorder. 
We find that these are the minimal ingredients necessary for the existence of our dissipative Bogoliubov-de Gennes (BdG) topological phase of amplification.

We present the following results:
\begin{itemize}
    \item The chain of coupled parametric oscillators has two topological phases with directional amplification, characterized by a winding number, $W_1$. In these phases, directional amplification is robust against disorder and imperfections.
    \item The first phase has $W_1 = 1$ and it is dominated by the parametric terms, which is why we name it dissipative BdG phase. It displays amplification in one quadrature of the field and squeezing in the other. The second topological phase has $W_1 = 2$ and requires collective dissipation. In this case both field quadratures are simultaneously amplified.
    \item Remarkably, we find that directional amplification can also take place in topologically trivial regimes, i.e., $W_1=0$. However, in those cases directional amplification is not robust and becomes suppressed by small amounts of disorder in the system.
    \item We characterize the gain, the noise-to-signal ratio and resilience to disorder of the topological phases. We find that they lead to amplifiers with high gain and broad bandwidth, which can also feature quantum-limited noise. 
    \item We study the stability of the system, showing that it is size-dependent, and find range of parameters where robust, topological amplifiers can operate while remaining stable.
    \item Finally, we have show that our model can be implemented with different experimental techniques like Floquet engineering with trapped ions or coupled resonators, or by exploiting non-linear terms in superconducting circuits~\cite{Ramos2022}.
\end{itemize}
The manuscript is organized as follows: In Section~\ref{sec:Model} we introduce the model and its description in the input-output formalism. 
In Section~\ref{sec:Topology} we characterize the topological properties of the array of parametric oscillators, its robustness to disorder and the stability of the topological phases. 
In Section~\ref{sec:Squeezing}, we characterize the gain, noise-to-signal ratio, and squeezing of the topological phases. In addition, we describe how combining a Green's function approach with decimation techniques allows us to find analytical expressions in the semi-infinite case, which perfectly capture the numerical results for finite systems~\cite{AGL-Decimation}.
In Section~\ref{sec:Experiments} we discuss some possible experimental implementations with state-of-the-art techniques.
Finally, in Section~\ref{sec:Conclusions}, we summarize our results and discuss future directions.
\section{Traveling wave-parametric amplifier: Model and Input-Output Theory\label{sec:Model}}
\subsection{Master equation}
The array of coupled parametric oscillators can be described by the master equation:
\begin{equation}
    \partial_t \rho = - i [ H, \rho] + {\cal L}_d( \rho ) + {\cal L}_p( \rho),
\end{equation}
with the first term describing the Hamiltonian dynamics, whereas the second and third terms describe incoherent losses and pump, respectively.

The Hermitian dynamics is described by the Hamiltonian:
\begin{eqnarray}
    H&=&\Delta\sum_{j=0}^{N-1}a_{j}^{\dagger}a_{j}+g_{s}\sum_{j=0}^{N-1}\left(a_{j}^{2}+a_{j}^{\dagger2}\right)\\ \label{eq:Hamiltonian}
    &&+\sum_{j=0}^{N-2}\left(Je^{i\phi}a_{j+1}^{\dagger}a_{j}+g_{c}a_{j}a_{j+1}+\text{h.c.}\right).\nonumber
\end{eqnarray}
There, the first line contains the local modes energy $\Delta$ and the single mode parametric terms $g_s$, while the second line contains the complex hopping between sites $Je^{\pm i\phi}$ and the two-mode parametric terms $g_c$.
This Hamiltonian can be implemented quite naturally in superconducting circuits~\cite{Blais2012}, without the need of Floquet or reservoir engineering, by just doing four-wave mixing in arrays of Josephson junctions and linear oscillators~\cite{Ramos2022}. 
There, the combination of Kerr effect and parametric driving can effectively provide the parametric terms present in Eq.~\eqref{eq:Hamiltonian}.

The dissipative losses in the master equation can be summarized by the following term:
\begin{equation}
    {\cal L}_d(\rho)=\sum_{j,l} \gamma_{j,l}{\cal D}[a_j,a_l^\dag](\rho),
\label{Ld}
\end{equation}
where ${\cal D}[A,B](\rho) = A\rho B-(BA\rho + \rho BA)/2$ and $\gamma_{j,l}$ describes collective decay processes between sites $j$ and $l$.
In this work we will consider local terms only, $\gamma_{j,l}=\gamma\delta_{j,l}$, which can naturally describe cavity losses in photonic setups or the coupling to a superconducting transmission line in arrays of Josephson junctions.

Similarly, the incoherent pump can be described by the term:
\begin{equation}
    {\cal L}_p(\rho)=\sum_{j,l} P_{j,l}{\cal D}[a_j^\dagger,a_l](\rho),
\end{equation}
In our particular case, we are interested in the contribution produced by coupling the system to auxiliary reservoirs shared by nearest neighbors. 
Tracing-out the reservoirs by adiabatic elimination, it produces the following contribution:
\begin{equation}
    P_{j,l}=2P \left( 2\delta_{j,l}+\delta_{j,l+1}+\delta_{j,l-1}\right),
\end{equation}
which includes a local term and a dissipative hopping, as indicated in Fig.~\ref{fig:Schematic} by green arrows. The local term naturally appears in the adiabatic elimination due to the possibility to hop back and forth between a site and the auxiliary reservoir~\cite{AGL-Keldysh,AGL-Decimation}.

As we will show below, the incoherent pump term is not required to engineer the dissipative BdG topological phase, but it is crucial to produce a topological phase with larger winding number. This means that a phase of topological amplification can be engineered with just local dissipative losses and parametric driving terms.
\subsection{Input-output theory\label{sec:Input-output}}

Although the master equation by itself would allow us to calculate many quantities of interest, it is very convenient to express the problem in the input-output formalism~\cite{Zoller-QuantumNoise,PhysRevA.103.033513}.
The quantum Langevin equation for the photonic modes can be written in the following form:
\begin{align}
     \partial_t\vec{a}(t)= -iH_{\rm nh}\vec{a}(t) +  \vec{\xi}_{\rm in}(t),\label{eq:Langevin}
\end{align}
where we have defined the Nambu spinors $\vec{a}(t)=[a_0(t),a_1(t),\ldots,a_0^\dag(t),a_1^\dag(t),\ldots]^{T}$ and the corresponding noise terms $\vec{\xi}_{\rm in}(t)=[\xi_0^{\rm in}(t),\xi_1^{\rm in}(t),\ldots,\xi_0^{\rm in}{}^\dag(t),\xi_1^{\rm in}{}^\dag(t),\ldots]^T$ (details regarding the noise terms are given in the Appendix~\ref{sec:Appendix-InputOutput}). 
The $2N\times2N$ non-Hermitian dynamical matrix has the following structure:
\begin{align}
    H_{\rm nh}=\begin{pmatrix}
    J+i\Gamma & K \\
    -K^\ast & -J^\ast+i\Gamma^\ast
    \end{pmatrix} ,
\label{eq:EffectiveH1} 
\end{align}
and each block is an $N\times N$ matrix with elements:
\begin{eqnarray}
    \Gamma_{j,l}&=&\left(\frac{4P-\gamma}{2}\right)\delta_{j,l}+P\left(\delta_{j,l+1}+\delta_{j,l-1}\right)\label{eq:coupling1},\\
    J_{j,l}&=&J\left(e^{-i\phi}\delta_{j,l+1}+e^{i\phi}\delta_{j,l-1}\right)\label{eq:coupling2},\\
    K_{j,l}&=&g_s\delta_{j,l}+g_c\left(\delta_{j,l+1}+\delta_{j,l-1}\right)\label{eq:coupling3}.
\end{eqnarray}
In the steady state, we can apply a Fourier transform to the operators, $a_j(\omega) = (2\pi)^{-1/2}\int dt e^{i\omega t} a_j(t)$, and write the solution to the system of equations as:
\begin{align}
    \vec{a}(\omega)=iG(\omega)\vec{\xi}_{\rm in}(\omega),\label{eq:FourierSpace}
\end{align}
where we have defined the dissipative Green's function~\cite{AGL-Decimation}:
\begin{equation}
G(\omega)=(\omega - H_{\rm nh})^{-1},\label{eq:G.w}
\end{equation}
and the Fourier transform of the Nambu spinors, $\vec{a}(\omega)=[a_0(\omega),a_1(\omega),\ldots,a_0^\dag(-\omega),a_1^\dag(-\omega),\ldots]^T$ and $\vec{\xi}_{\text{in}}(\omega)=[\xi^{\rm in}_0(\omega),\xi^{\rm in}_1(\omega),\ldots,\xi_0^{\rm in}{}^\dag(-\omega),\ldots]^T$.
Finally, from Eq.~(\ref{eq:FourierSpace}) we can write the explicit solution for the Fourier transform of the operators:
\begin{align}
    a_j (\omega) = i\sum_{l=0}^{N-1} \left[ G_{j,l}(\omega)\xi_l^{\rm in}(\omega)+G_{j,N+l}(\omega)\xi_l^{\rm in}{}^\dag(-\omega)\right] \label{linearsolution}
\end{align}
Then, from Eq.~(\ref{linearsolution}), the solution for $G(\omega)$ and the input-output relation:
\begin{equation}
    a^{\rm out}_j(\omega) = a^{\rm in}_j(\omega)+\sqrt{\gamma}a_j(\omega)\label{eq:input-output},
\end{equation}
we can calculate arbitrary correlation functions of the output field at some particular site, $j$.

For the characterization of the amplification properties we are interested in the propagation of an input signal inserted at the edge and detected at a particular site, $j$.
For example, if the input signal is given by a coherent state with amplitude $\alpha$ and frequency $\omega_d$, we can write the output field as the average value and its fluctuations:
\begin{equation}
    a_j^{\text{out}}(\omega)=\langle a_j^{\text{out}}(\omega) \rangle + \delta a_j^{\text{out}}(\omega),
\end{equation}
where the average value can be calculated using Eq.~\eqref{eq:input-output}, to give:
\begin{align}
    \langle a_{j}^{\text{out}}\left(\omega\right)\rangle=&\alpha\delta\left(\omega-\omega_{d}\right)\left[\delta_{j,0}-i\gamma G_{j,0}\left(\omega\right)\right]\nonumber\\
    &-i\alpha^{*}\delta\left(\omega+\omega_{d}\right)\gamma G_{j,N}\left(\omega\right) .
\end{align}
Notice how the presence of parametric driving produces an output signal at two different frequencies, $\pm\omega_d$, typically referred to as signal and idler, respectively.
From this, we can define the gain of the amplifier at site $j\neq0$ as:
\begin{equation}
    \mathcal{G}_j(\omega)=\gamma^2 |G_{j,0}(\omega)|^2 .
\label{eq:Gain}
\end{equation}
Analogously, as we are also interested in the noise properties of the amplifier, we define the normalized noise-to-signal ratio:
\begin{equation}
    n^\text{add}_j(\omega)=\frac{n_{j}^{\text{amp}}(\omega)}{\mathcal{G}_{j}(\omega)},\label{eq:signal-noise-ratio}
\end{equation}
with $n_{j}^{\text{amp}}(\omega)$ being the noise added by the amplifier:
\begin{align}
    n^{\text{amp}}_j (\omega)=&\gamma^2 \sum_{l=0}^{N-1}|G_{j,N+l}\left(\omega\right)|^{2}\nonumber\\
    &+\gamma \sum_{l,l^{\prime}=0}^{N-1}P_{l^{\prime},l}G_{j,l}^{*}\left(\omega\right)G_{j,l^{\prime}}\left(\omega\right).
\end{align}
Finally, we are also interested in the possibility of generating squeezed states. 
For their characterization we define the following Fourier transform of the quadrature operators for the output fields:
\begin{align}
    X_{j}^{\text{out}}\left(\omega,\theta\right)=& a_{j}^{\text{out}}\left(\omega\right)e^{i\theta}+a_{j}^{\text{out}\dagger}\left(-\omega\right)e^{-i\theta},\\
    P_{j}^{\text{out}}\left(\omega,\theta\right)=& ia_{j}^{\text{out}}\left(\omega\right)e^{i\theta}-ia_{j}^{\text{out}\dagger}\left(-\omega\right)e^{-i\theta},\label{eq:Quadratures}
\end{align}
where $\theta$ is an angle that determines along which direction the quadratures are measured.
From the input-output relations and the solution to the photonic modes in terms of Green's functions, we can express the variance of a quadrature,
\begin{eqnarray}
    \Delta\mathcal{O}_j(\omega)=\sqrt{\langle \mathcal{O}_j^{\text{out}}(\omega) \mathcal{O}_j^{\text{out}}(-\omega) \rangle},
\end{eqnarray}
in terms of the input modes, being $\mathcal{O}^{\text{out}}_j=X^{\text{out}}_j$ or $P^{\text{out}}_j$. Since the Heisenberg uncertainty corresponds to $\Delta X \Delta P\geq1$, a variance below $1$ implies that the corresponding quadrature is squeezed.
\section{Topological phases and amplification \label{sec:Topology}}
\begin{figure*}
    \centering
    \includegraphics[width=1\textwidth]{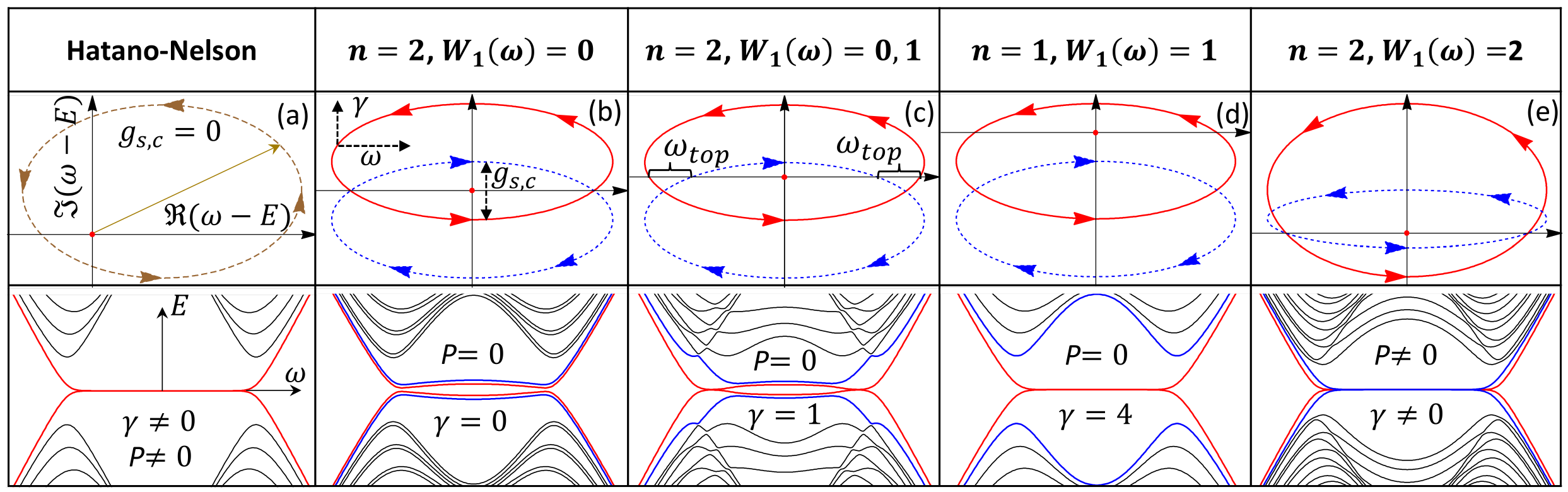}
    \caption{Comparison between different amplification phases with $n$ zero-energy modes and topological invariant, $W_1(\omega)$. The spectrum of the doubled matrix with OBC is calculated for a uniform distribution of on-site disorder, $\Delta\to\Delta+w$, averaged over $100$ realizations and with $w/J\in[-0.2,0.2]$. For PBC, loops enclosing the origin lead to $W_{\pm}=\pm 1$. As indicated in (b), $\gamma$ shifts both loops in the vertical axis, while $\omega$ shifts them along the horizontal one. In contrast, $g_{s,c}$ splits the loops along the vertical axis in opposite directions. The different columns correspond to: (a) the Hatano-Nelson model with $g_{s,c}=0$; (b) the trivial phase of the TWPA with $\kappa=P=0$ and two pairs of degenerate zero-energy modes split by disorder; (c) the BdG phase with $\gamma/J=1$, which removes the splitting of the zero-energy for $\omega\in\omega_\text{top}$, making them topological and resilient to disorder; (d) the same phase with $\gamma/J=4$, where now a single pair of degenerate and topological zero-energy modes is always present; and (e) the double Hatano-Nelson phase with $P=0.75$ and $\gamma/J=4$, where as both winding vectors rotate in the same direction, the two pairs of edge states become topological. In absence of disorder the spectra in (b), (c) and (d) would display full degeneracy between all zero-energy modes.}
    \label{fig:Table}
\end{figure*}
\subsection{General theory of topological amplification}
Recent theoretical work has revealed a connection between the theory of Hermitian topological insulators and directional amplification in gain/loss systems~\cite{PhysRevLett.122.143901,PhysRevA.103.033513,AGL-Keldysh}.
In this section we review this theoretical framework and explain how it can be extended to predict robust topological phases of amplification in bosonic BdG arrays with dissipation.
We will show that in this model, the presence of zero-energy modes leads to directional amplification which is not always topologically protected, and that one of the topological phases displays common features with the Kitaev chain in the absence of time-reversal symmetry and its $\mathbb{Z}_{2}$ invariant~\cite{Kitaev_Majorana,Thesis-Majoranas}.

We are interested in the topological characterization of the Green's function defined in Eq.~\eqref{eq:G.w}. 
For that, we first define the Hermitian matrix:
\begin{eqnarray}
\mathcal{H}\left(\omega\right)&=&
\left(\begin{array}{cc}
0 & \omega-H_{\text{nh}}\\
\omega-H_{\text{nh}}^{\dagger} & 0
\end{array}\right)\label{eq:DoubledMatrix},
\end{eqnarray}
which we will refer to as \textit{doubled matrix}. 
The rationale for defining $\mathcal{H}\left(\omega\right)$ is two-fold. First, the Green's function, $G(\omega)$, can be directly written in terms of its eigenstates. Second, as we show below, the existence of zero-energy eigenstates of $\mathcal{H}\left(\omega\right)$ imply directional amplification along the system.

To proceed, if $\tau_z$ acts on the degree of freedom produced by doubling the Hilbert space in $\mathcal{H}(\omega)$, notice that the doubled matrix has an intrinsic chiral symmetry:
\begin{eqnarray}
\tau_z \mathcal{H}\left(\omega\right) \tau_z = - \mathcal{H}\left(\omega\right),
\label{eq:chiral}
\end{eqnarray}
which is an inherent mathematical property of $\mathcal{H}\left(\omega\right)$, independent of the physical symmetries of the underlying lattice. 
Kramer's theorem implies that eigenstates come in pairs:
\begin{equation}
    {\cal H}(\omega) 
    \left( 
    \begin{array}{c}
    \vec{u}_n \\
    \pm \vec{v}_n 
    \end{array}
    \right) = 
    \pm E_n 
        \left( 
    \begin{array}{c}
    \vec{u}_n \\
    \pm \vec{v}_n 
    \end{array}
    \right),
\end{equation}
where $\vec{u}_n$, $\vec{v}_n$, $n = 0, \dots, N-1$, 
are $N$ normalized vectors, and the eigenvalues, $E_n$, are positive numbers. 
Actually,  $\vec{u}_n$, $\vec{v}_n$ and $E_n$ are also the singular vectors and singular values of $H_{\textrm{nh}}$, respectively, and one can write $\omega-H_{\rm nh} = \sum_n \vec{u}_n E_n \vec{v}_n^{\dagger}$.
For that reason, the Green's function can be written as:
\begin{equation}
    G(\omega) = \sum_n \vec{v}_n \frac{1}{E_n} \vec{u}^{\dagger}_n ,
\label{eq:svd}
\end{equation}
This implies that, if $\mathcal{H}(\omega)$ is in a topologically non-trivial phase, we expect the appearance of zero-energy modes and an amplification effect in Eq.~\eqref{eq:svd}.\\
To see this, consider the simplest case of a single zero-energy mode. 
Due to the mapping to an Hermitian Hamiltonian in Eq.~\eqref{eq:DoubledMatrix}, the topological phase will display a pair of zero-energy modes localized at the boundary. 
In a finite system, their splitting will be exponentially suppressed with the length of the chain $E_0 \propto e^{-N\zeta}$, being $\zeta$ the inverse correlation length, and this mode will be separated from the rest of eigenstates of 
${\cal H}(\omega)$ by a finite gap.
As such, the sum in Eq.~\eqref{eq:svd} will be dominated by this term and we can approximate the sum over $n$ by:
\begin{equation}
G(\omega) \propto e^{N\zeta} \vec{v}_0 \vec{u}^{\dagger}_0,
\end{equation}
leading to a few important conclusions:
(i) the response of the system is amplified as a function of the inverse correlation length, (ii) the spatial distribution of the Green's function reflects the spatial distribution of the zero-energy mode of the doubled matrix, which is typically localized a the edges of the chain, and (iii) the amplification of the Green's function is topologically protected against disorder that preserves the symmetries of the topological phase.

Let us now particularize the analysis to our current dissipative BdG system with Periodic Boundary Conditions (PBC):
\begin{equation}
    H_{\text{nh}}(k)= f_0(k)\mathbf{1}+\vec{f}(k)\cdot\vec{\sigma},\label{eq:H_nh}
\end{equation}
where we have written the dynamical matrix, Eq.~\eqref{eq:EffectiveH1}, in terms of the Pauli matrices $\vec{\sigma}=(\sigma_x,\sigma_y,\sigma_z)$ acting in the Nambu subspace, and the coefficients $f_j(k)$:
\begin{align}
    f_{0}\left(k\right)=&-2J\sin\left(k\right)\sin\left(\phi\right)\\\nonumber
    &-i\frac{\gamma}{2}+4iP\cos^{2}\left(\frac{k}{2}\right), \\
    f_{x}\left(k\right)=&0,\ f_{y}\left(k\right)=i\left[g_{s}+2g_{c}\cos (k)\right], \\
    f_{z}\left(k\right)=&\Delta+2J\cos (k)\cos(\phi) .
\end{align}
Notice that in addition to the dissipative terms, which are proportional to $\gamma$ and $P$, the parametric terms, proportional to $g_{s,c}$, also break Hermiticity. This is a common feature in bosonic BdG Hamiltonians, consequence of the Bogoliubov transformation used to diagonalize the system, which has to conserve bosonic statistics~\cite{BdG-Diagonalization-Colpa1986,BdG-Brandes2015}.

As the doubled matrix is Hermitian by construction, we can now study its topology in terms of the standard classification of topological insulators~\cite{Ryu_2010,TI3}.
Interestingly, the presence of the intrinsic chiral symmetry, $\tau_z$, means that our topological classification only requires to consider chiral symmetric classes, in which Particle-Hole Symmetry (PHS) and Time-Reversal Symmetry (TRS) are both simultaneously present or absent.
The symmetry analysis reveals that the system is in the AIII class, which lacks TRS and PHS, and is characterized by a $\mathbb{Z}$ invariant in 1D (details in the Appendix~\ref{sec:Appendix-Topology}).
Only for $\phi=0 \mod(\pi)$ the class changes to CI, which is trivial in 1D.

From all these considerations, we conclude that \textit{the system can display topological properties, if and only if, TRS is broken via the phase $\phi$}, and that the relevant topological invariant is given by~\cite{TI3}:
\begin{equation}
    W_{1}(\omega)=\int_{-\pi}^{\pi}\frac{dk}{4\pi i}\textrm{tr}\left[\tau_z\mathcal{H}(k,\omega)^{-1}\partial_k\mathcal{H}(k,\omega)\right],\label{eq:Top-invariant}
\end{equation}
with $\mathcal{H}(k,\omega)$ being the Fourier transform of the doubled matrix. Then, by virtue of Eq.~\eqref{eq:svd}, $W_1(\omega) \neq 0$ will imply the existence of directional amplification with topological protection, i.e., with resilience to disorder that conserves the symmetries of the topological class. Importantly, as the AIII class only has the chiral symmetry, which is present by construction in $\mathcal{H}(\omega)$, we can predict strong resilience to all types of disorder.
%%%%%%%%%%

Our theory of topological amplification can be illustrated with a simple example based on the Hatano-Nelson model (full details can be found in our previous work~\cite{PhysRevA.103.033513}).
This is actually a limiting case of our model, if we remove the parametric terms (i.e., if we set $g_{c,s} = 0$), which leads to $K = 0$ in Eq.~\eqref{eq:EffectiveH1}. 
In that situation, the use of Nambu spinors is not needed and $H_{\text{nh}}$ is a $N \times N$ non-Hermitian matrix, whose representation in momentum space is a scalar function of $k$, $H_{\rm nh}(k)$.\\
In the Hatano-Nelson model, the evaluation of the winding number from Eq.~\eqref{eq:Top-invariant}, results in $W_1 = 1$ when the loop formed by the vector $\vec{h}(\omega,k)=\left( \Re \left[\omega- E(k)\right], \Im \left[\omega- E(k)\right] \right) $, encloses the origin, being $E(k)=H_\text{nh}(k)$ the eigenvalue in this scalar case. 
This is represented in Fig.~\ref{fig:Table}~(a), where the top row shows the trajectory of $\vec{h}(\omega,k)$ for a particular case in the topological phase, and the bottom row shows the spectrum of $\mathcal{H}(\omega)$ for a finite system, in the presence of disorder, confirming its robustness due to the absence of splitting between the degenerate pair of zero-energy modes. 

Coming back to our model, we find that in contrast with the Hatano-Nelson model, $H_{\text{nh}}(k)$ from Eq.~\eqref{eq:H_nh} is now a $2 \times 2$ matrix. 
However, the geometrical interpretation of $W_1(\omega)$ in terms of a loop in the complex plane can be maintained, although the extra dimension from the BdG structure now requires to consider two loops instead.
To show this, we re-write $\mathcal{H}(k,\omega)$ in Eq.~\eqref{eq:Top-invariant} in terms of $H_{\rm nh}(k)$.
Then after some manipulation, it can be written as (details in the Appendix~\ref{sec:Appendix-Topology}):
\begin{eqnarray}
W_1(\omega)=W_{+}(\omega)+W_{-}(\omega),
\end{eqnarray}
where $W_{\pm}(\omega)$ are the winding numbers for the two eigenvalues of $H_{\rm nh}(k)$:
\begin{equation}
    W_{\pm}(\omega)=\frac{1}{2\pi i}\int_{-\pi}^{\pi} \partial_k \log\left[\omega-E_{\pm}(k)\right]dk,
\end{equation}
which take values:
\begin{align}
    E_{\pm}(k)=&f_0(k)\pm \sqrt{\left[f_z(k)\right]^2+\left[ f_y(k) \right]^2}.
\end{align}
As for models containing up to nearest neighbors contributions, $W_{\pm}(\omega)=0$, $1$, we can conclude that the winding number will take values $W_1(\omega) = 0$, $1$, $2$, leading to the same number of topologically protected zero-energy states (see Fig.~\ref{fig:Table} for a summary of the different cases). 
Remarkably, we will show below that, in addition to the topologically protected zero-energy states, the system can also host zero-energy states which produce exponential amplification, but are not topologically protected, as shown in Fig.~\ref{fig:Table}~(b).
%%%%
\subsection{Dissipative BdG topological phase ($P=0$)}
%%%%
We first assume the presence of local dissipation only by setting $P=0$.
The implementation in this case is simpler, as it does not require external incoherent pump and all the dissipative processes are local.
\begin{figure}
    \centering
    \includegraphics[width=1\columnwidth]{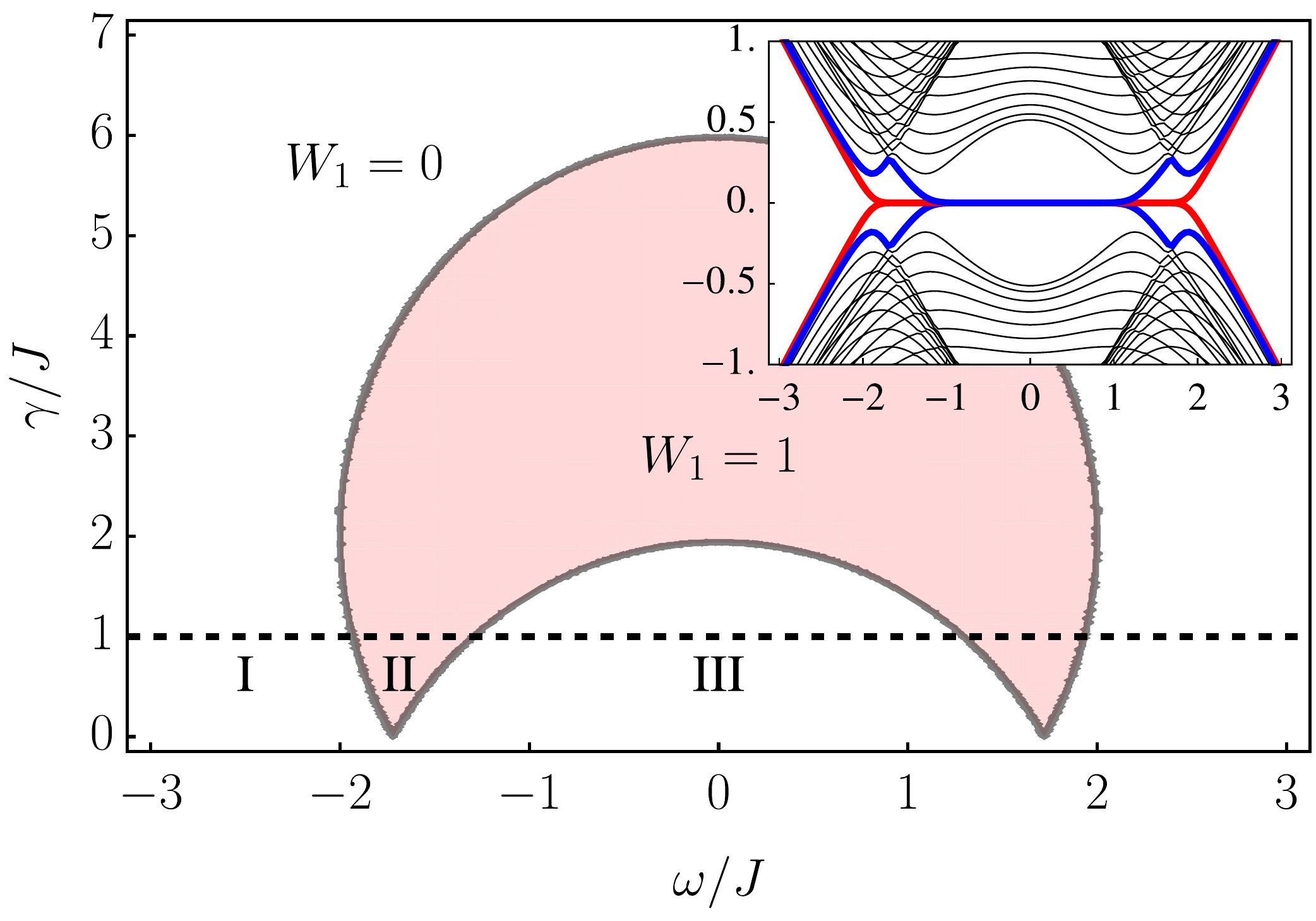}
    \caption{Topological phase diagram as a function of $\gamma/J$ and $\omega/J$ for $g_{s,c}/J=1$, $\phi=\pi/2$ and $\Delta/J=0$. The inset shows the eigenvalues of $\mathcal{H}(\omega)$ for the cut $\gamma/J=1$, indicated by a dashed line. Clearly, one can differentiate three regions: Region II is topological with a degenerate pair of zero-energy modes, while regions I and III are topologically trivial but differ in their number of zero-energy modes. The spectrum is plotted for a finite system with $N=50$.}
    \label{fig:Fig-Z2-PhaseDiag}
\end{figure}
To study the topology we calculate the value of the winding number from Eq.~\eqref{eq:Top-invariant}, as a function of frequency and losses.
This is a natural choice because $\gamma$ can be easily controlled in Josephson arrays by just changing the coupling to an auxiliary transmission line, and plotting vs $\omega$ provides useful information about the frequency bandwidth of the amplifier.
The topological phase diagram is shown in Fig.~\ref{fig:Fig-Z2-PhaseDiag}, for $g_{s}/J=g_{c}/J=1$ and $\phi=\pi/2$. 

In general, $g_{s,c}$ needs to be non-vanishing to obtain a topological phase, but their ratio only affects the shape of the topological region. 
Similarly, as previously discussed in the symmetry analysis, $\phi$ needs to be different from $0$ or $\pi$, which is why we fixed its value $\phi=\pi/2$.
Fig.~\ref{fig:Fig-Z2-PhaseDiag} shows that a non-trivial topological phase requires dissipation, and that for $2<\gamma<6$, a wide range of frequencies display topological amplification.

We plot in the inset of Fig.~\ref{fig:Fig-Z2-PhaseDiag} the spectrum of $\mathcal{H}(\omega)$ for an array with $N=50$ sites and $\gamma/J=1$, indicated by a dashed line in the figure.
Interestingly, one can see three different regions as a function of $\omega$, labeled as I, II and III. Region I corresponds to the standard trivial region, lacking of zero-energy modes. Region II corresponds to a topological phase with $W_1(\omega)= 1$ and a pair of degenerate zero-energy modes (indicated in red). Finally, region III is not topological because $W_1(\omega)=0$, but it has two pairs of degenerate zero-energy modes (indicated in red and blue).\\
The unexpected presence of region III is interesting, as it indicates that exponential amplification can also be present for $W_1(\omega)=0$. This is something that has been overlooked in previous works, and shows that the existence of exponential amplification does not imply the existence of non-trivial topological phases with $W_1\neq0$.

\begin{figure}
    \centering
    \includegraphics[width=1\columnwidth]{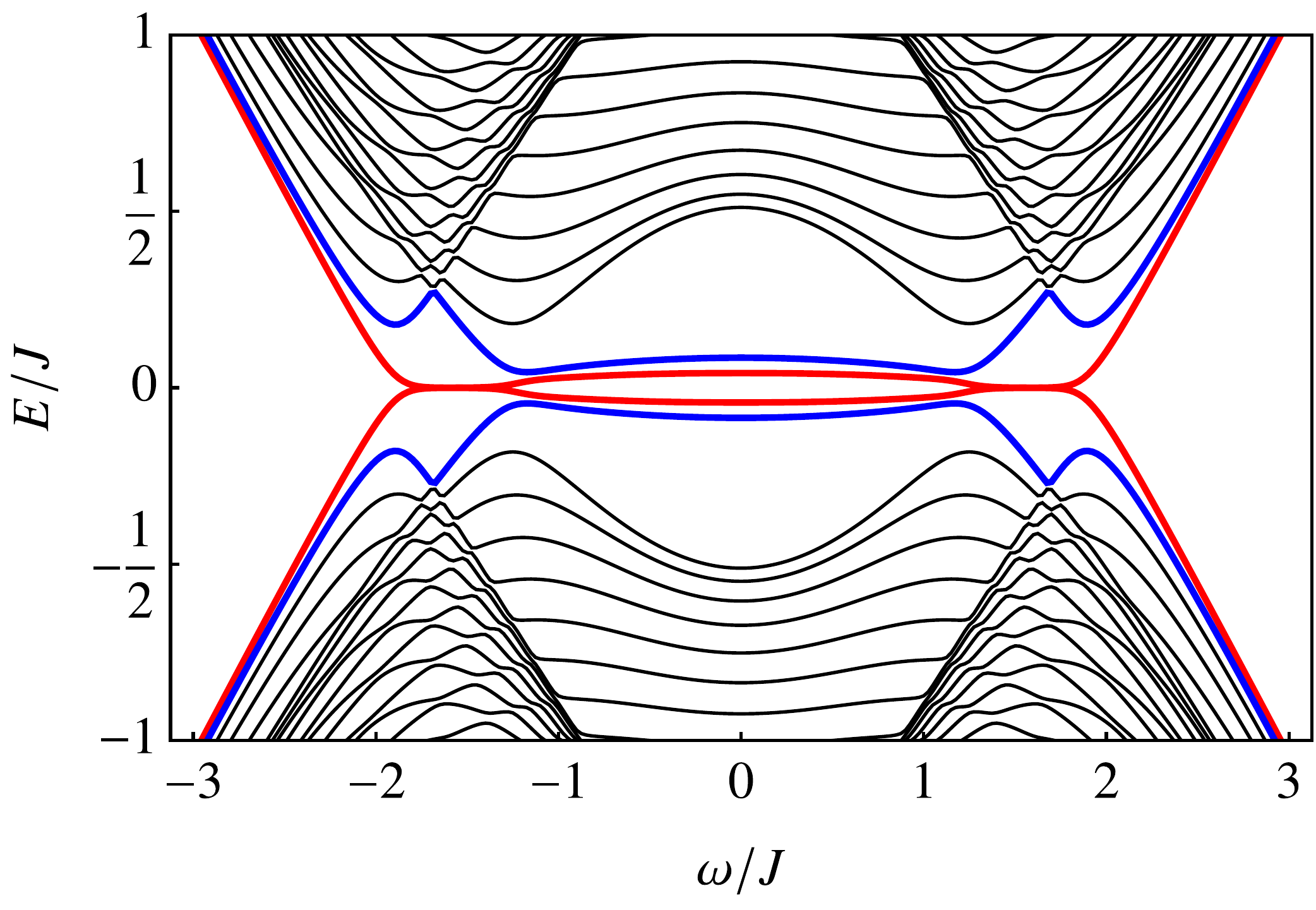}
    \caption{Eigenvalues of $\mathcal{H}(\omega)$ for $g_{c,s}/J=1$, $\phi=\pi/2$, $\gamma/J=1$, $\Delta=w$ and $N=50$, for a uniform disorder distribution with $w/J\in [-0.2,0.2]$, averaged over $100$ realizations. In red/blue are highlighted the zero-energy modes which split in the region where they were degenerate. However, in the region with a single pair (red), the zero-energy modes remain pinned to zero.}
    \label{fig:RobustnessZ2}
\end{figure}

To illustrate the practical importance of this feature, in Fig.~\ref{fig:RobustnessZ2} we characterize the robustness to disorder of the zero-energy modes.
This plot shows the same spectrum as the inset of Fig.~\ref{fig:Fig-Z2-PhaseDiag},
however, on-site disorder $w/J\in[-0.2,0.2]$ has now been added, leading to a splitting in region III where the zero-energy modes were doubly degenerate. 
In contrast, the topological region with a single pair of zero-energy modes remains unaffected, confirming its topological robustness to perturbations.
This effect is reminiscent to what is observed in other BdG Hamiltonians, such as in the case of Majorana fermions in the Kitaev chain~\cite{Kitaev_Majorana}. There, in the presence of TRS breaking terms, the system is in the D class and its topology is characterized by a $\mathbb{Z}_2$ invariant. Then, when two Majorana fermions occupy the same site, they are no longer topologically protected and the splitting produced by disorder is linearly suppressed, rather than exponentially~\cite{Thesis-Majoranas}.
Hence, only edges with an odd number of Majorana zero energy modes remain topologically protected.
This case is summarized in columns (b), (c) and (d) of Fig.~\ref{fig:Table} and geometrically explained in terms of the addition of winding numbers, $W_{\pm}(\omega)$.

\begin{figure}
    \centering
    \includegraphics[width=1\columnwidth]{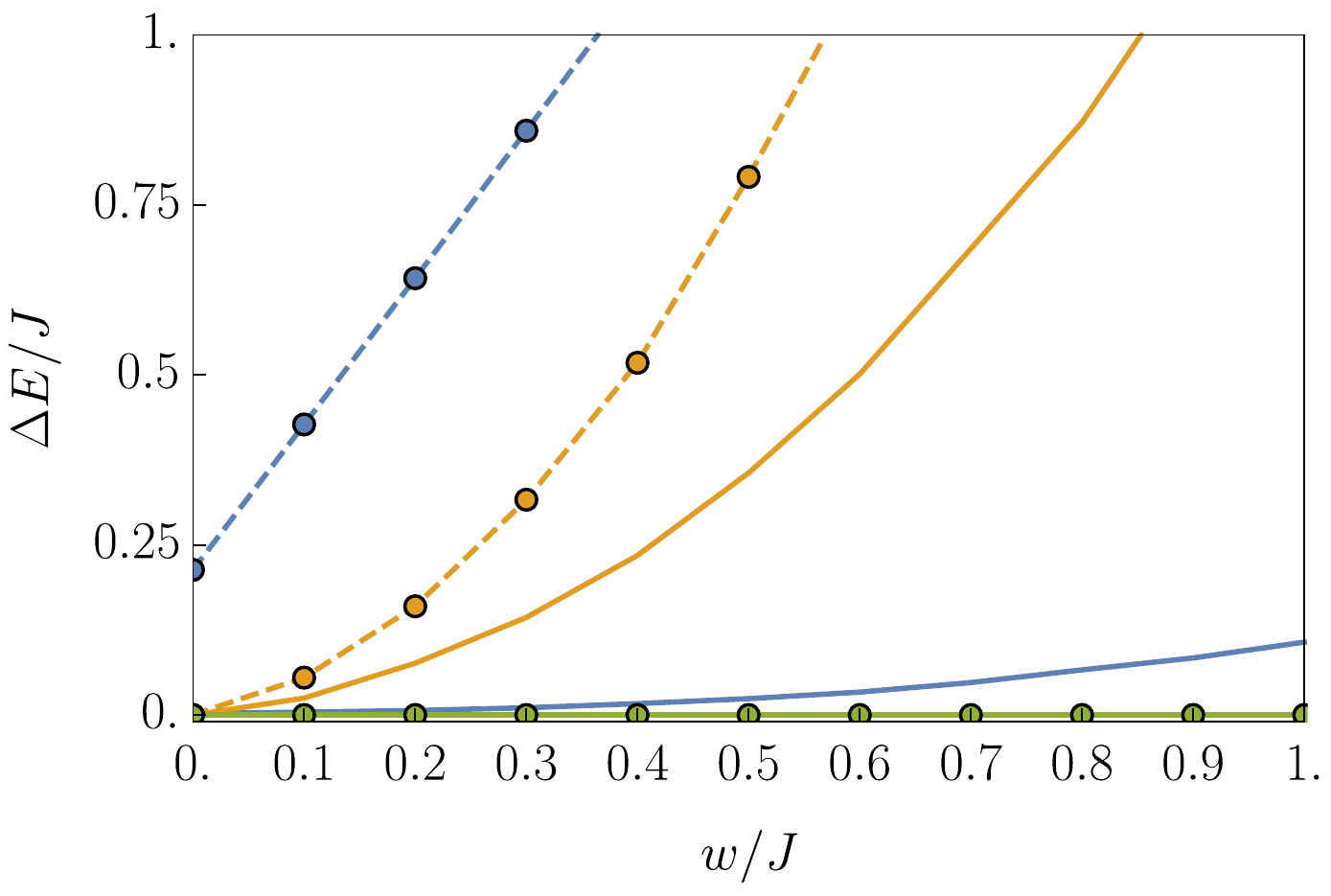}
    \caption{Energy of the two pairs of modes with smallest energy as a function of a uniform distribution of on-site disorder, $w/J$, averaged over $100$ realizations. The solid line corresponds to the lowest energy pair of zero-energy modes and the dotted one to the next pair in energy. The colors yellow, blue and green correspond to cases shown in Fig.~\ref{fig:Table}, (b), (d) and (e), respectively.}
    \label{fig:Splitting}
\end{figure}
To quantify the resilience to disorder, we plot in Fig.~\ref{fig:Splitting} the energy of the two pairs of modes with smallest energy in $\mathcal{H}(\omega)$, as a function of the disorder strength, $w/J$. 
The solid line indicates the splitting from zero of the lowest energy pair of states, while the dotted line indicates the splitting of the next pair in increasing energy. The yellow color corresponds to the case $\gamma/J=0$, shown in Fig.~\ref{fig:Table}~(b), where the system is in a trivial phase with two pairs of degenerate zero-energy modes. Increasing disorder immediately splits the two pairs from zero energy, confirming their lack of topological protection.
In contrast, the blue color represents the case $\gamma/J=4$, shown in Fig.~\ref{fig:Table}~(d), where just a single pair of zero-energy modes is present, but it is topologically protected. For this reason, the zero-energy modes~(solid line), require to go beyond $w/J>0.2$ to separate from $E=0$. In addition, the dotted line shows that the next pair of states in energy are not zero-energy modes, and respond linearly to disorder.
Finally, the green lines corresponds to a different topological phase, which will be discussed in the following sections.\\
Importantly, notice that robustness to disorder is expected in all parameters and not just in the local energies. This is explicitly demonstrated in Ref.~\cite{Ramos2022}, where an experimental proposal of this phase using Josephson junctions also tests the resilience of amplification and noise to all types of disorder that could be present in the sample.

These results show that the presence of exponential amplification with the distance is not in one-to-one correspondence with topology, as it can be produced in a topologically trivial phase.
Therefore, if the system is in the topological phase of Fig.~\ref{fig:Table}~(d), it will display robustness to disorder, indicating that the performance of the amplifier should not be greatly affected by details of the fabrication process. 
In contrast, if the system is in the trivial phase shown in Fig.~\ref{fig:Table}~(b), it will suffer a continuous decline of its amplifying properties, although it could still be used as an amplifier.
From the intermediate case, such as that of Fig.~\ref{fig:Table}~(c) , we can deduce that the role of local loss $\gamma$, is to break the degeneracy between pairs of zero-energy modes. 
This removes one pair of zero-energy modes, but makes the remaining one topologically protected to disorder.

This means that for the experimental implementation of the dissipative BdG topological phase of amplification, one must try to increase dissipative losses until all frequencies are topologically amplified (e.g., to values $\gamma/J\geq 2$, as shown in Fig.~\ref{fig:Fig-Z2-PhaseDiag}).
However, as we discuss next, to produce a reliable topological amplifier the stability of the topological phase must also be taken into account.
\subsubsection{Stability of the dissipative BdG topological phase}
\begin{figure}
    \centering
    \includegraphics[width=1\columnwidth]{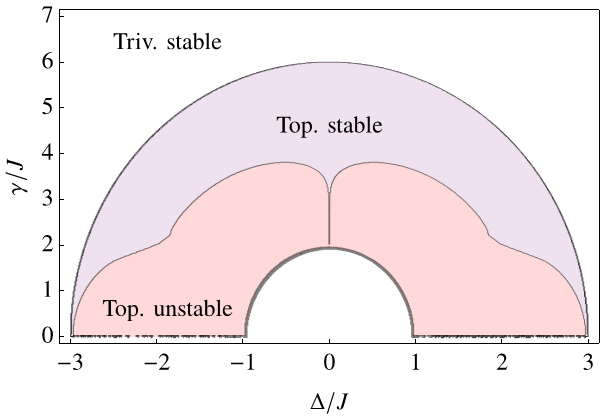}
    \includegraphics[width=1\columnwidth]{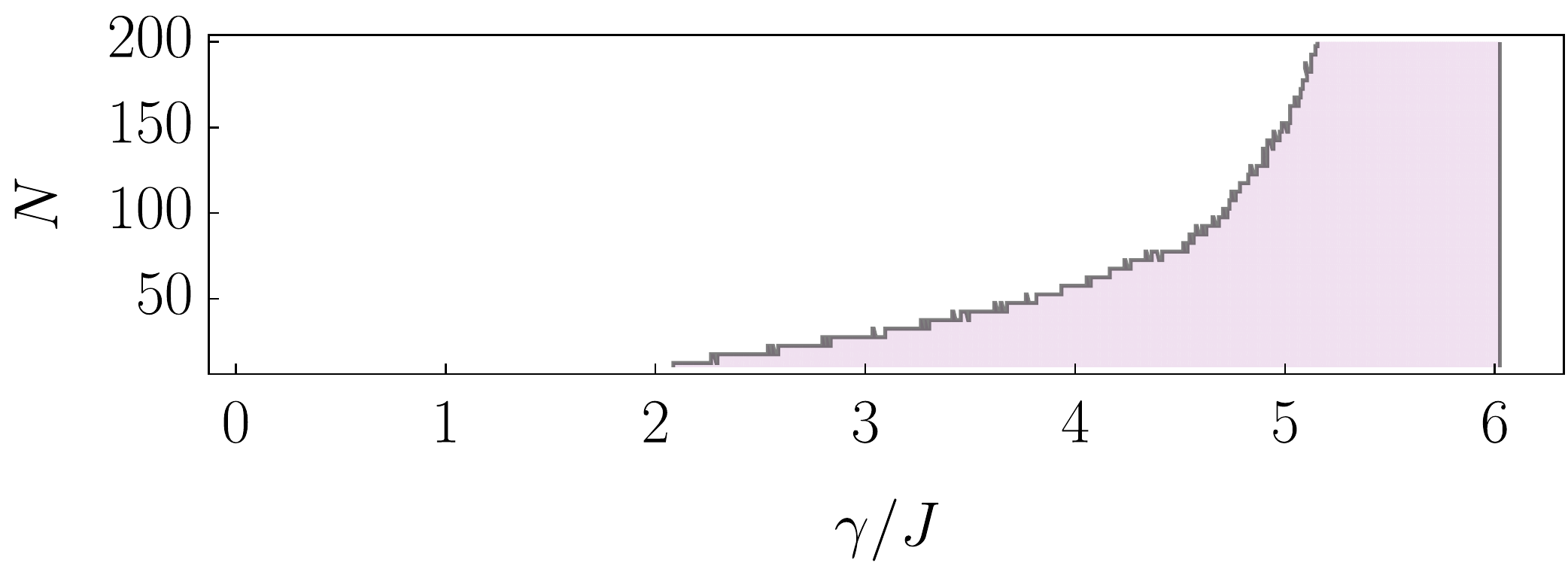}
    \caption{Top: Stability diagram as a function of local dissipation $\gamma/J$ and detuning $\Delta/J$ for $N=12$, $\phi=\pi/2$ and $g_{s,c}/J=1$. The topological phases from Fig.~\ref{fig:Fig-Z2-PhaseDiag} divide into stable (purple) and unstable (red) regions. Bottom: Stable regions of the topological phase as a function of $\gamma/J$ and $N$, for $\Delta/J=0$.}
    \label{fig:Stability1}
\end{figure}
The stability of the amplifier can be characterized in terms of the eigenvalues of the dynamical matrix $H_{\text{nh}}$, for open boundary conditions.
If their imaginary part becomes positive, the system becomes unstable and its physical realization is not feasible.
This relation between complex eigenvalues and stability is specially interesting in the case of dissipative topological phases, where the skin effect can drastically change the spectrum~\cite{PhysRevLett.121.086803,PhysRevLett.124.086801}. As point-gap topological phases with periodic boundary conditions always have a positive imaginary part, they are unstable. However, for open boundary conditions the skin effect changes the spectrum and they can become stable.
An intuitive way to see this consists in imagining the differences between a signal traveling in a loop or in a finite line. In the former case the signal is amplified indefinitely, while the later amplifies the signal just over a finite length.\\
In Fig.~\ref{fig:Stability1}~(top), we combine the topological phase diagram for the winding number with the stability of the corresponding regions, as a function of $\Delta/J$ and $\gamma/J$, for an array with $N=12$ sites.
It shows that the topological phase separates into topologically stable and topologically unstable parts.
In the Hatano-Nelson model, these two regions have been shown to display different dynamical properties, related with the steady state of the system~\cite{PhysRevB.95.064302,AGL-Decimation}.
However, in contrast with the Hatano-Nelson model, we find that stability in this model is size-dependent. 
We have numerically checked that this dependence appears due to the presence of parametric terms, as the limit $g_{s,c}/J=0$ removes this dependence. 
Curiously, we also find that the trivial phase with two pairs of degenerate zero-energy modes is unstable. This is not a general feature, as one can stabilize this phase for $g_s<g_c$. However, Fig.~\ref{fig:Stability1} shows that the role of $\gamma> 0$ in the realization of the topological amplifier is not just to produce $W_1(\omega)\neq0$, but also to stabilize the system.
Therefore, local losses not only split the degenerate pairs of zero-energy modes and produce a topologically protected phase of amplification, they also stabilize the topological amplifier.\\
To study the size-dependence of stability in more detail, we plot in Fig.~\ref{fig:Stability1}~(bottom), the region of stability within the topological phase, as a function of dissipation and the size of the array. 
We find that the stable region slowly shrinks as the size increases, however, as the system is characterized by exponential amplification, arrays with a small number of sites are reasonable choices and their region of stability is large.

In conclusion, a stable topological phase requires a balance between $g_{s,c}/J$, $\gamma/J$ and $N$.
This is because $g_{s,c}/J$ can be reduced to keep the system stable for larger $N$ and produce zero-energy modes close to the thermodynamic limit. 
However, this also reduces the gap between the zero-energy modes and the bulk states, making the topological phase more fragile to disorder.
Despite these limitations, in Section~\ref{sec:Squeezing} we will show that good amplifiers can be produced within these constrains. However, let us first describe the topology and stability of the topological phase in the presence of collective pump.
\subsection{Double Hatano-Nelson phase ($P \neq 0$)}
Previously, we neglected collective dissipative terms, but their role in topological amplification can be important. 
This is the case in the Hatano-Nelson model, where non-local pump is a necessary ingredient to find non-trivial topology~\cite{PhysRevA.103.033513}. 
Similarly, collective dissipative terms in this model can directly affect the values that the winding number can take. 
For example, we saw that for $P/J=0$, the winding number can only take values $W_1(\omega)=0,1$, even in the presence of two pairs of degenerate zero-energy modes in the gap. The reason is that $E_{\pm}(k)$ will always wind in opposite directions and the corresponding winding numbers, $W_\pm(\omega)$, cancel, if they simultaneously are not zero.
Therefore, it would be interesting to find a way to make all the degenerate pairs of zero-energy modes from Fig.~\ref{fig:Fig-Z2-PhaseDiag} topologically protected against disorder.
Below we show that for $P\neq0$, $W_1(\omega)$ can reach larger values and that all its zero-energy modes are topologically protected.
\begin{figure}
    \centering
    \includegraphics[width=1\columnwidth]{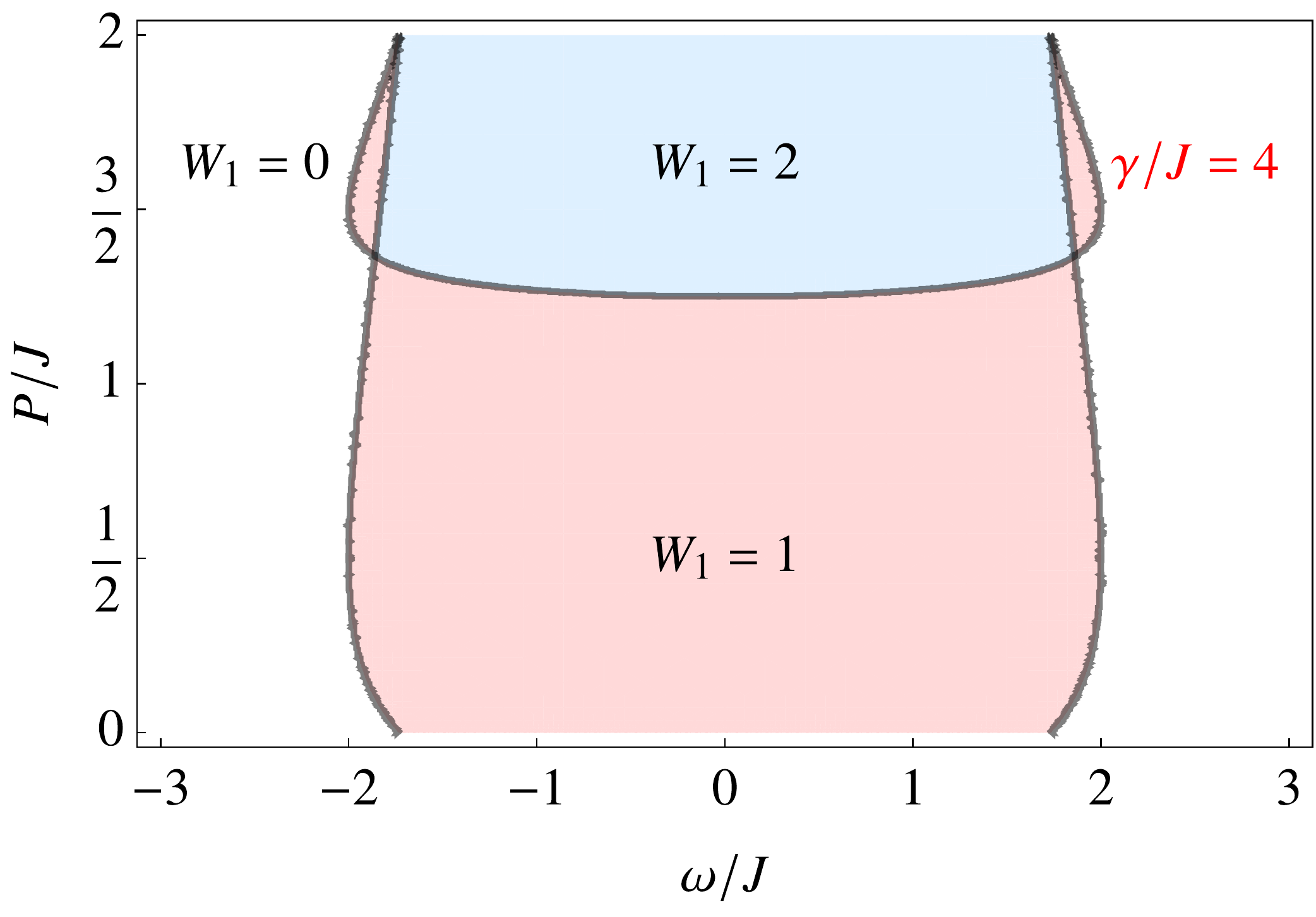}
    \caption{Topological phase diagram as a function of $P/J$ and $\omega/J$. Parameters: $g_{s,c}/J=1$, $\Delta/J=0$, $\phi=\pi/2$ and $\gamma/J=4$.}
    \label{fig:PhaseDiagramZ-0}
\end{figure}

In Fig.~\ref{fig:PhaseDiagramZ-0} we plot the topological phase diagram, as a function of $P/J$, for the case of $\gamma/J=4$. At this value of $\gamma/J$, the system initially is in a topological phase with $W_1(\omega)=1$ and has a single degenerate pair of zero-energy modes in the gap (see Fig.~\ref{fig:Fig-Z2-PhaseDiag}). One can see that the presence of collective pump eventually leads to $W_1(\omega)=2$, when $P/J$ is sufficiently large.
The reason for this change in the winding number can be intuitively understood from the geometrical picture in Fig.~\ref{fig:Table}~(e). There, one can see that the $k$-dependence from the collective pump changes the loop direction, initially controlled by $g_{c}$. Hence, if $P \gtrsim g_{c}$, the two winding numbers $W_{\pm}(\omega)$ rotate in the same direction and do not cancel.
We must remark that the value of $P/J$ does not affect the topological class of $\mathcal{H}(k,\omega)$, because only $\phi$ controls this change, as previously argued.
A closely related case has been analyzed in the context of the non-Hermitian Skin effect~\cite{PhysRevLett.124.086801}.

\begin{figure}
    \centering
    \includegraphics[width=1\columnwidth]{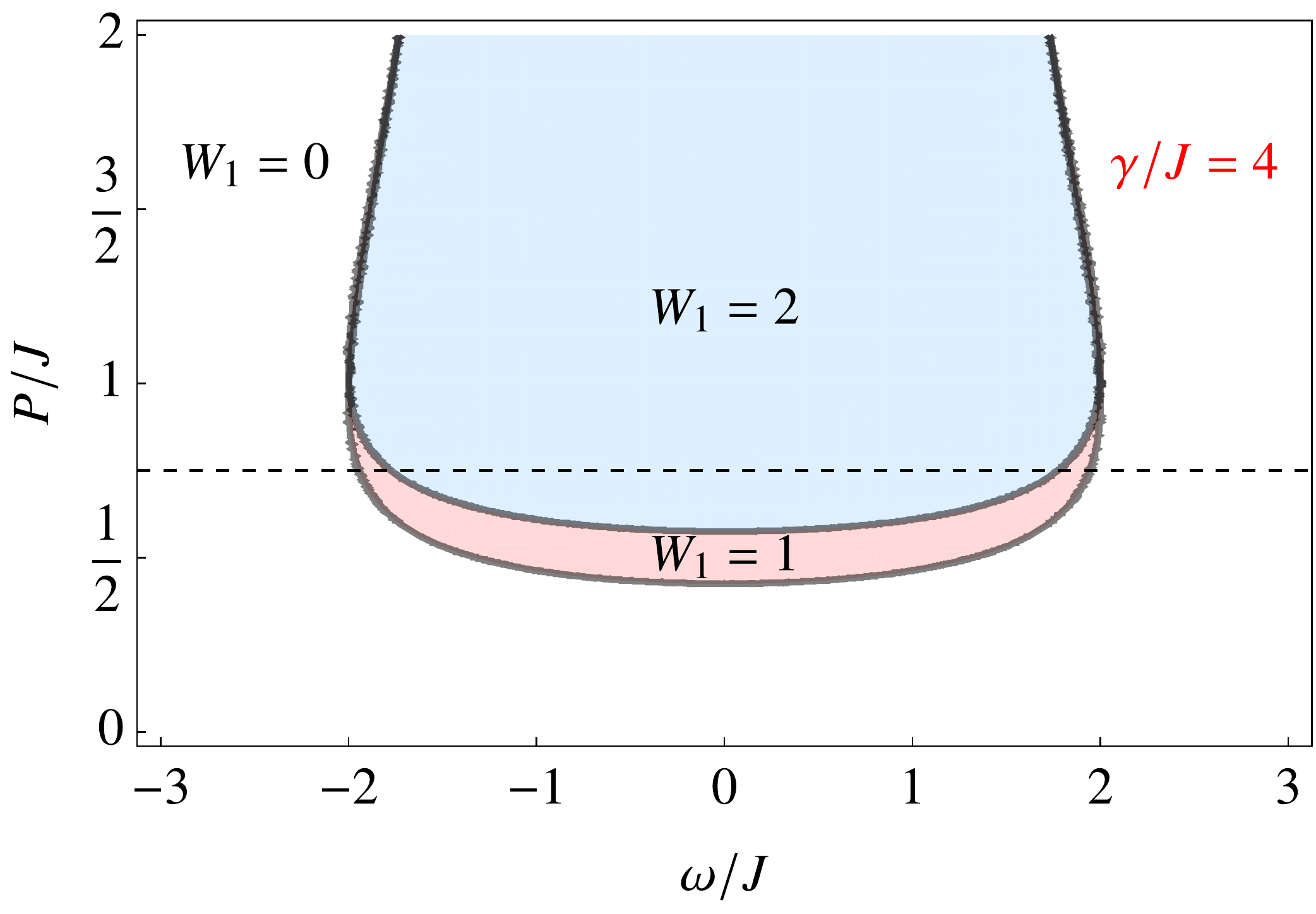}
    \caption{Topological phase diagram as a function of $P/J$. Parameters: $g_{s,c}/J=0.1$, $\Delta/J=0$, $\phi=\pi/2$ and $\gamma/J=4$. We have reduced the value of $g_{s,c}/J$ for stability reasons discussed in the main text. This would only re-scale the phase diagram of Fig.~\ref{fig:Fig-Z2-PhaseDiag}.}
    \label{fig:PhaseDiagramZ}
\end{figure}

Physically, this regime is dominated by incoherent pump and loss, and reducing the value of $g_{s,c}/J$ does not qualitatively affect the phase with $W_1(\omega)=2$. Their decrease mainly shrinks the area where the phase with $W_1(\omega)=1$ exists (cf Fig.~\ref{fig:PhaseDiagramZ-0} and Fig.~\ref{fig:PhaseDiagramZ}, where the only difference is a decrease in $g_{s,c}/J$ from $1$ to $0.1$).
Actually, this topological phase has important similarities with the one in the Hatano-Nelson model. It can be checked that $g_{s,c}/J$ can be reduced to an arbitrarily small value without changing the winding number from $W_1(\omega)=2$, and that in this limit, the model reduces to the Hatano-Nelson. This means that the topological phase can be understood as two weakly coupled copies of the Hatano-Nelson model, one for positive frequency and other for negative frequency modes.
In particular for this phase, the sole effect of the parametric terms will be to make the propagation for positive and negative frequency modes slightly different, as we will show in the next section.
However, it is important to address first the stability of this topological phase. 
\subsubsection{Stability of the double Hatano-Nelson phase}
It turns out that, because of the presence of collective pump, the stability of the phase is severely affected.
This can be intuitively understood from the fact that both, collective pump and parametric terms destabilize the system, and this must be compensated by increasing losses. However, too large losses can drive the system into a trivial phase.
Fortunately, we know that one of the sources of instability is the presence of parametric terms, which also turns stability into a size-dependent property.
As in addition we are aware that it is possible to reduce their value without changing the invariant, we will use this feature to find a stable phase with $W_1(\omega)=2$.

In Fig.~\ref{fig:PhaseDiagramZ} we plot the topological phase diagram, now for the case of $g_{s,c}/J=0.1$. As previously anticipated, the region with $W_1(\omega)=1$ is reduced, but the one with $W_1(\omega)=2$ exists for a wide range of parameters.
It is important to mention that in this topological phase with smaller $g_{s,c}/J$, resilience to disorder remains strong. Naively, one could think that this would not be the case, because the gap should be reduced by decreasing the squeezing terms. However, this is compensated by the contribution from collective pump, $P/J$. This can be confirmed in Fig.~\ref{fig:Splitting} (green), where it is shown that the degeneracy of all the zero-energy modes (solid and dotted line, which completely overlap) is robust in the presence of disorder, and that their resilience is present up to very large values.
\begin{figure}
    \centering
    \includegraphics[width=1\columnwidth]{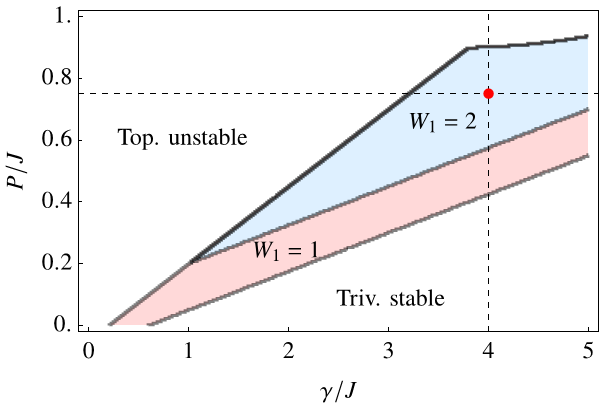}
    \caption{Stability of the topological phase as a function of the dissipative parameters. Parameters: $\omega/J=0$, $g_{s,c}/J=0.1$,$\Delta/J=0$, $\phi=\pi/2$ and $N=12$. The red dot indicates the point chosen to plot the spectrum in Fig.~\ref{fig:RobustnessZ} and to calculate the amplifier properties in Sec.~\ref{sec:Squeezing}.}
    \label{fig:Stability2}
\end{figure}

\begin{figure}
    \centering
    \includegraphics[width=1\columnwidth]{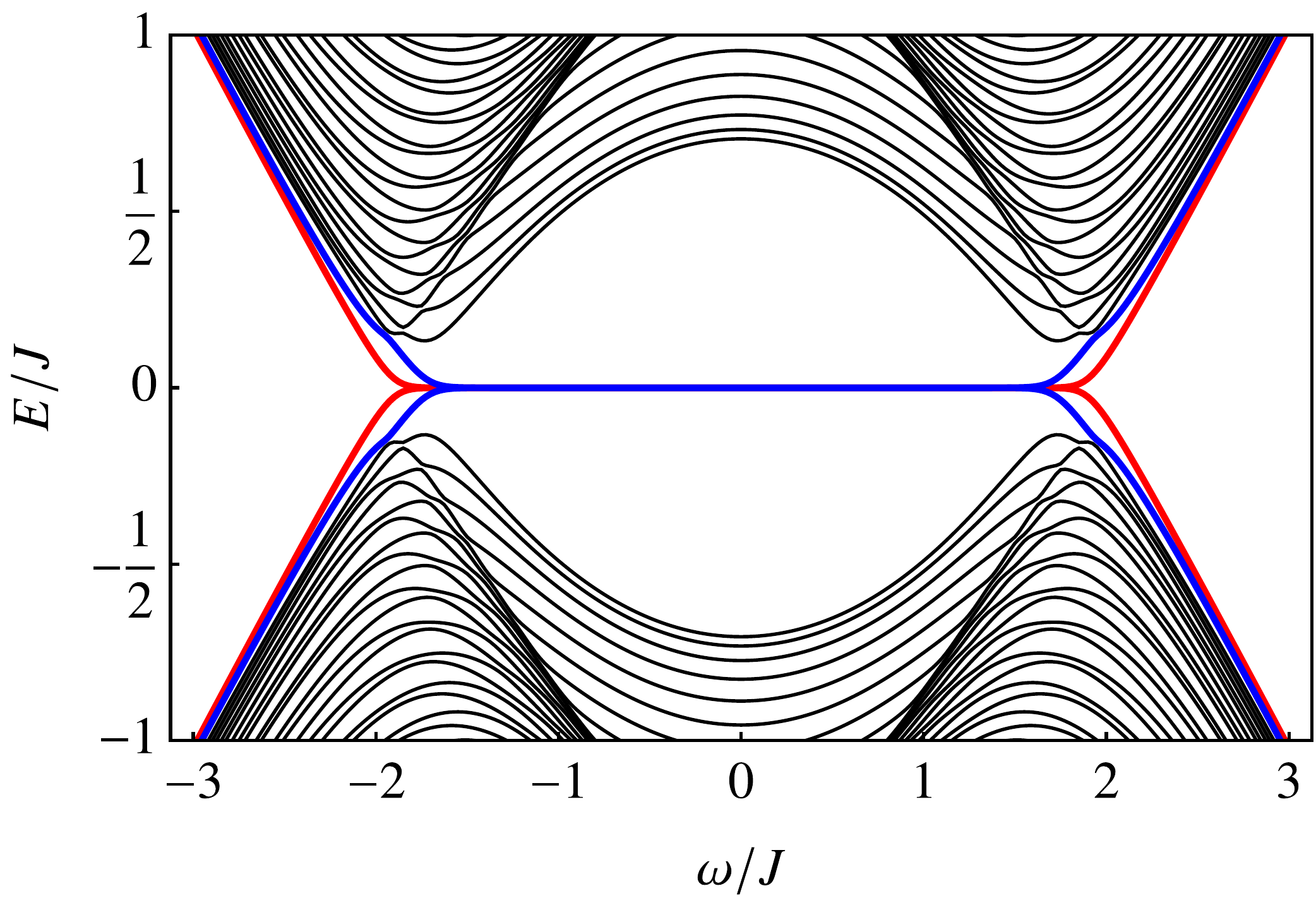}
    \caption{Eigenvalues of $\mathcal{H}(\omega)$ for $N=50$, $g_{c,s}/J=0.1$, $\phi=\pi/2$, $\gamma/J=4$, $P/J=0.75$, $\Delta/J=0$ and and uniform disorder distribution $w/J\in [-0.2,0.2]$, averaged over $100$ realizations. In red/blue are highlighted the zero-energy modes, which remain robust to disorder.}
    \label{fig:RobustnessZ}
\end{figure}

To confirm that for small parametric terms, $g_{s,c}$, the topological phase with $W_1(\omega)$ can be stabilized, we plot in Fig.~\ref{fig:Stability2} the stable topological phases as a function of the dissipative parameters, for a system with $N=12$ sites.
It shows that a stable phase with $W_1(\omega)=2$ is possible, and in addition, the stability of the topological phases as a function of the size improves, because decreasing $g_{s,c}/J$ reduces the size-dependence of the stable regions. For example, the case $P/J=0.75$ with $\gamma/J=4$, indicated by a red dot, remains stable up to a size of $N\simeq50$.
However, notice that a large part of the phase diagram in Fig.~\ref{fig:PhaseDiagramZ} remains unstable.
To confirm the resilience to disorder, we plot in Fig.~\ref{fig:RobustnessZ} the spectrum of $\mathcal{H}(\omega)$ at this point, for a system with $N=50$ sites and a uniform distribution of on-site disorder averaged $100$ times. This confirms that the phase with $W_1(\omega)=2$ has two pairs of degenerate zero-energy modes, and that they are topologically protected, even for moderate disorder.

In summary, we have found that the addition of collective pump terms allows to find a topological phase of amplification with a larger number of topological zero-energy modes, although its stability requires to reduce the squeezing terms.
Besides its intrinsic interest regarding the topology of BdG systems, this phase is also interesting from the perspective of topological amplification, because it allows us, for the first time, to explore phases with more than one channel of amplification.
We will discuss in the next section the details of the amplification in each phase and show that they can be very different.
From now on, the red dot in Fig.~\ref{fig:Stability2} indicates the parameters chosen to study this stable phase with $W_1(\omega)=2$.
\section{Amplifier properties: gain, noise and squeezing\label{sec:Squeezing}}
We now explore the physical properties of the different topological phases using a Green's function approach.
This is very adequate for topological amplification in dissipative systems, as it has been shown that, in combination with decimation, it is possible to obtain accurate analytical approximations in the semi-infinite size limit of 1D systems~\cite{PhysRevB.95.064302}. This limit is specially appealing to study topological features, as it connects the bulk properties of the array, with the presence of topological edge states due to the boundary.

The calculation requires to first determine the surface Green's function, $\hat{G}_{0,0}\left(\omega\right)$, from the non-linear matrix equation~\cite{AGL-Decimation}:
\begin{equation}
    \hat{G}_{0,0}\left(\omega\right)=\hat{g}_{0,0}\left(\omega\right)\left[1+\mathcal{V}_{+}\hat{G}_{0,0}\left(\omega\right)\mathcal{V}_{-}\hat{G}_{0,0}\left(\omega\right)\right].\label{eq:non-linearEq}
\end{equation}
Here, the hat indicates that the surface Green's functions are $2\times2$ matrices with components:
\begin{eqnarray}
    \hat{G}_{j,l}\left(\omega\right)=\left(\begin{array}{cc}
G_{j,l}\left(\omega\right) & G_{j,N+l}\left(\omega\right)\\
G_{N+j,l}\left(\omega\right) & G_{N+j,N+l}\left(\omega\right)
\end{array}\right)
\end{eqnarray}
written in the Nambu spinor basis $\vec{a}_j(\omega)=[a_j(\omega),a_j^{\dagger}(-\omega)]^T$.
This form is very convenient for practical calculations, as these are the four components coupled by the parametric terms.
Therefore, the dissipative Green's function for an isolated site is given by:
\begin{eqnarray}
    \hat{g}_{0,0}(\omega)=\left(\begin{array}{cc}
    \omega-\Delta+i\frac{\gamma-4P}{2} & -g_{s}\\
    g_{s} & \omega+\Delta+i\frac{\gamma-4P}{2}
    \end{array}\right)^{-1}.\label{eq:Unperturbed-g00}
\end{eqnarray} 
In addition, the hopping matrices, $\mathcal{V}_{\pm}$, describe the forward and backward complex hopping between sites, respectively:\begin{eqnarray}
    \mathcal{V}_{\pm}&=\left(\begin{array}{cc}
    Je^{\pm i\phi}+iP & g_{c}\\
    -g_{c} & -Je^{\mp i\phi}+iP
    \end{array}\right).\label{eq:Hopping-matrix}
\end{eqnarray}
In some cases, an analytical solution for the matrix $\hat{G}_{0,0}(\omega)$ is not possible. However, this is unimportant, as we can always find a numerical solution and use it to calculate the physical observables in the semi-infinite limit.
In this work, we will consider both scenarios: we will impose additional constraints between parameters to find analytical expressions, and also use numerical solutions to calculate the observables in a range of parameters where compact analytical expressions cannot be obtained. 
Simultaneously, we will compare our results in the semi-infinite case with the ones from exact diagonalization for a finite system, to check the accuracy and the importance of finite-size effects.

An important property of the surface Green's function is that it can be related with an arbitrary Green's function using~\cite{AGL-Decimation}:
\begin{widetext}
\begin{equation}
    \hat{G}_{j,l}(\omega)=\left[\hat{G}_{0,0}(\omega)\mathcal{V}_{\text{sgn}\left(l-j\right)}\right]^{\left|j-l\right|}\hat{G}_{0,0}(\omega)+\sum_{r=0}^{r_f}\left[\hat{G}_{0,0}(\omega)\mathcal{V}_{-}\right]^{j-r}\left[\hat{G}_{0,0}(\omega)\mathcal{V}_{+}\right]^{l-r}\hat{G}_{0,0}(\omega)\label{eq:GreenFunction1} ,
\end{equation}
\end{widetext}
where $r_f=\min\left\{j,l\right\} -1$.
Also, to relate the Green's function to the physical observables, it is useful to define an inverse coherence length, $\zeta_{\pm}(\omega)$.
A natural way to do this is to consider the propagation of an excitation from the edge to site $j$, which is characterized by the following Green's function:
\begin{equation}
    \hat{G}_{j,0}(\omega)=\left[\hat{G}_{0,0}(\omega)\mathcal{V}_{-}\right]^{j}\hat{G}_{0,0}(\omega) .
\label{eq:Propagator-Edge}
\end{equation}
If we rewrite this expression using the spectral decomposition of $G_{0,0}(\omega)\mathcal{V}_{-}$ in terms of its eigenvalues and projectors, $\lambda_{\pm}(\omega)$ and $P_{\pm}(\omega)$, respectively:
\begin{equation}
    \hat{G}_{0,0}(\omega)\mathcal{V}_{-}=\sum_{\alpha=\pm}\lambda_{\alpha}(\omega)P_\alpha(\omega),\label{eq:Projectors-decomp}
\end{equation}
we can write the following expression for the Green's function:
\begin{equation}
    \hat{G}_{j,0}(\omega)=\sum_{\alpha=\pm}e^{\zeta_{\alpha}(\omega)j}P_{\alpha}(\omega) \hat{G}_{0,0}(\omega) ,
\label{eq:propagator1}
\end{equation}
where we have defined the inverse coherence length as $\zeta_{\pm}(\omega)=\log[\lambda_{\pm}(\omega)]$ (details of the decomposition in the Appendix~\ref{sec:App-Decimation}).\\
Eq.~\eqref{eq:propagator1} has a useful structure, as it separates the local contribution from the one that depends on the relative distance between sites, $j$.
In physical terms, Eq.~\eqref{eq:propagator1} characterizes the propagation of an initial excitation along two orthogonal subspaces with different coherence lengths.
The condition $\Re[\zeta_{\pm}(\omega)]>0$, can be used to indicate the presence of amplification along each subspace, and $\Im[\zeta_{\pm}(\omega)]$ is the phase gained during propagation.
In addition, we will show below that the form of the projectors determines the mixing of positive and negative frequency modes in each subspace, and will be useful to characterize the squeezing generated in the system.

Now we calculate the surface Green's function for the different topological phases described in Section~\ref{sec:Topology}, and use this result to determine the properties of the amplifier.
\subsection{Dissipative BdG topological phase ($P=0$)}
\subsubsection{Inverse coherence length}
To determine $\zeta_{\pm}(\omega)$, we focus on the topological phase diagram of Fig.~\ref{fig:Fig-Z2-PhaseDiag}, where $g_{s,c}/J=1$, $\Delta/J=0$ and $\phi=\pi/2$, and determine its surface Green's function from Eq.~\eqref{eq:non-linearEq}.
In this case, there is a unique compact solution which can be used to write the decomposition of $G_{0,0}(\omega)\mathcal{V}_{-}$ in Eq.~\eqref{eq:Projectors-decomp}. We find the following eigenvalues and projectors:
\begin{equation}
    \lambda_{\pm}=\left\{ \frac{2iJ}{\omega-iJ+i\frac{\gamma}{2}},0\right\},\ P_{\pm}=\frac{1}{2}\left(\begin{array}{cc}
1 & \mp i\\
\pm i & 1
\end{array}\right) .
\label{eq:Projectors}
\end{equation}
The presence of the $\lambda_{-}=0$ eigenvalue indicates that the matrix $G_{0,0}(\omega)\mathcal{V}_{-}$ is singular, and that the physics can be reduced to that of the $P_{+}$ subspace, which mixes positive and negative frequency modes.
This is a particular feature for the present case with $g_{s}=g_{c}$. However, we generally find that this topological phase always has $\Re [\zeta_{-}(\omega)]\leq 0$ and will never produce amplification in the subspace defined by $P_-$.
That is, for $g_{s}\neq g_{c}$ and the system in the dissipative BdG topological phase, only one of the eigenvalues will have $\Re [\zeta_{+}(\omega)]>0$, while the other will always have $\Re [\zeta_{-}(\omega)]\leq 0$.
We show below how this is related with squeezing one quadrature while amplifying the other, and that this is not possible in the case with $W_1(\omega)=2$, because in both subspaces there is amplification.
\begin{figure}
    \centering
    \includegraphics[width=1\columnwidth]{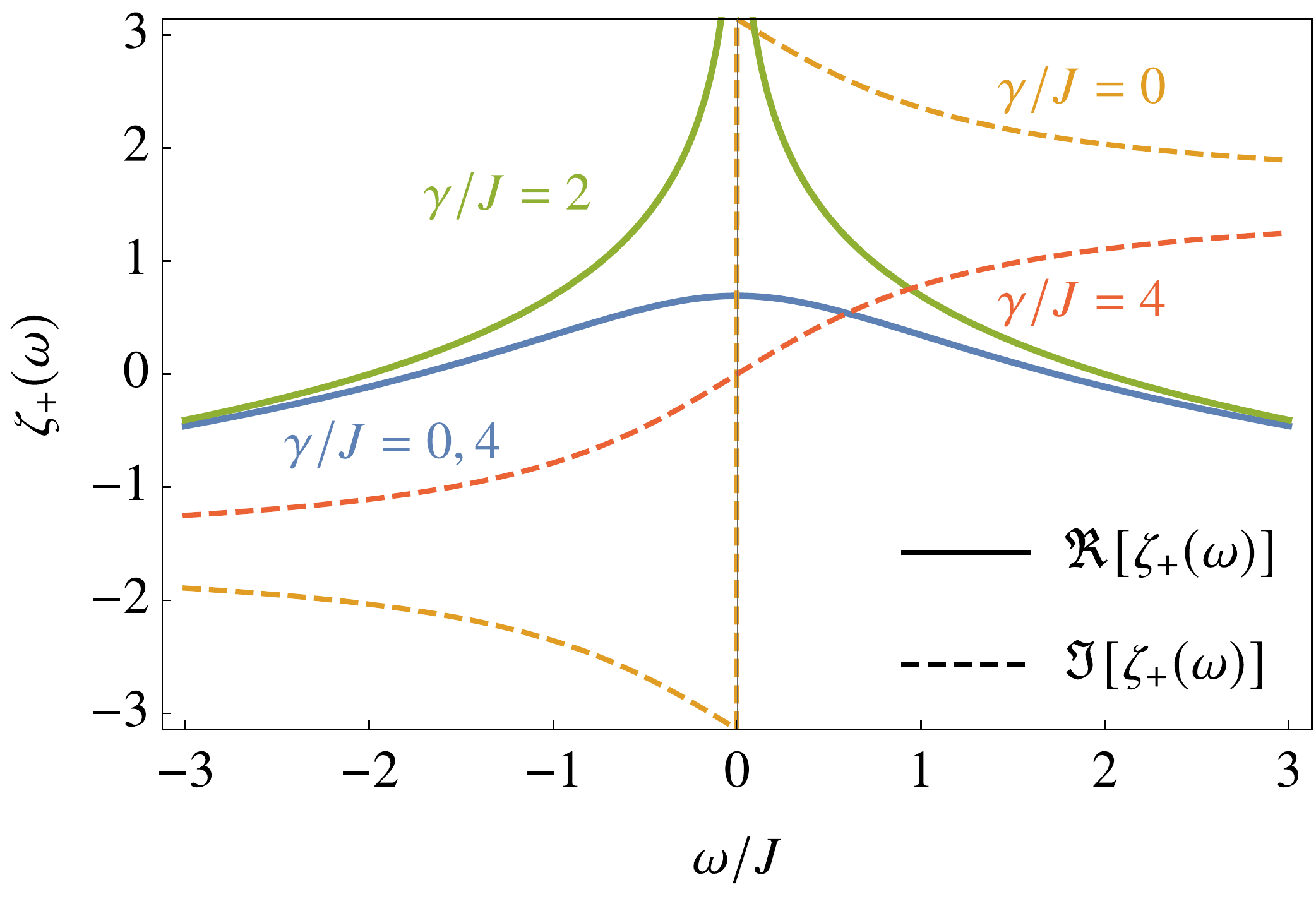}
    \caption{Real (solid) and imaginary part (dashed) of $\zeta_{+}(\omega)$ for different values of $\gamma/J$. The comparison with the one obtained from the exact diagonalization of a system with $N\gtrsim 10$ is excellent. In the range of frequencies with $\Re [\zeta_{+}(\omega)]>0$ signals are amplified.}
    \label{fig:CoherenceLength1}
\end{figure}

In Fig.~\ref{fig:CoherenceLength1} we plot the inverse correlation length as a function of frequency and for different values of losses. 
It shows that for the case $\gamma/J=0$~(blue), where two pairs of non-topological zero-energy modes coexist, amplification happens for the region with $\Re [\zeta_{+}(\omega)]>0$, but it is not topologically protected, as discussed in Section~\ref{sec:Topology}.
Interestingly, the case $\gamma/J=4$~(blue) shows an identical amplification profile, with the difference that in this case there is a single pair of topologically protected zero-energy modes.
This means that just by looking at the amplification of a signal, it would be difficult to differentiate between topological and non-topological amplification. 
Only by looking at their robustness to disorder or by measuring the phase acquired during propagation given by $\Im[\zeta_\pm(\omega)]$ in Fig.~\ref{fig:CoherenceLength1} (dashed yellow and red), would it be possible to detect the topological origin of amplification.

An additional interesting feature in Fig.~\ref{fig:CoherenceLength1} is the divergence of $\zeta_{+}(\omega)$ for $\omega/J=0$ and $\gamma/J=2$~(green), where the system becomes completely directional. This is the only point where $\lambda_{+}$ can diverge and coincides with the value of $\gamma/J$ where the zero-energy modes become topological for the whole range of frequencies where they exist. 
Hence, it is similar to a critical point in Quantum Phase Transitions, where correlations diverge.
\subsubsection{Amplifier gain and noise}

\begin{figure}
    \centering
    \includegraphics[width=1\columnwidth]{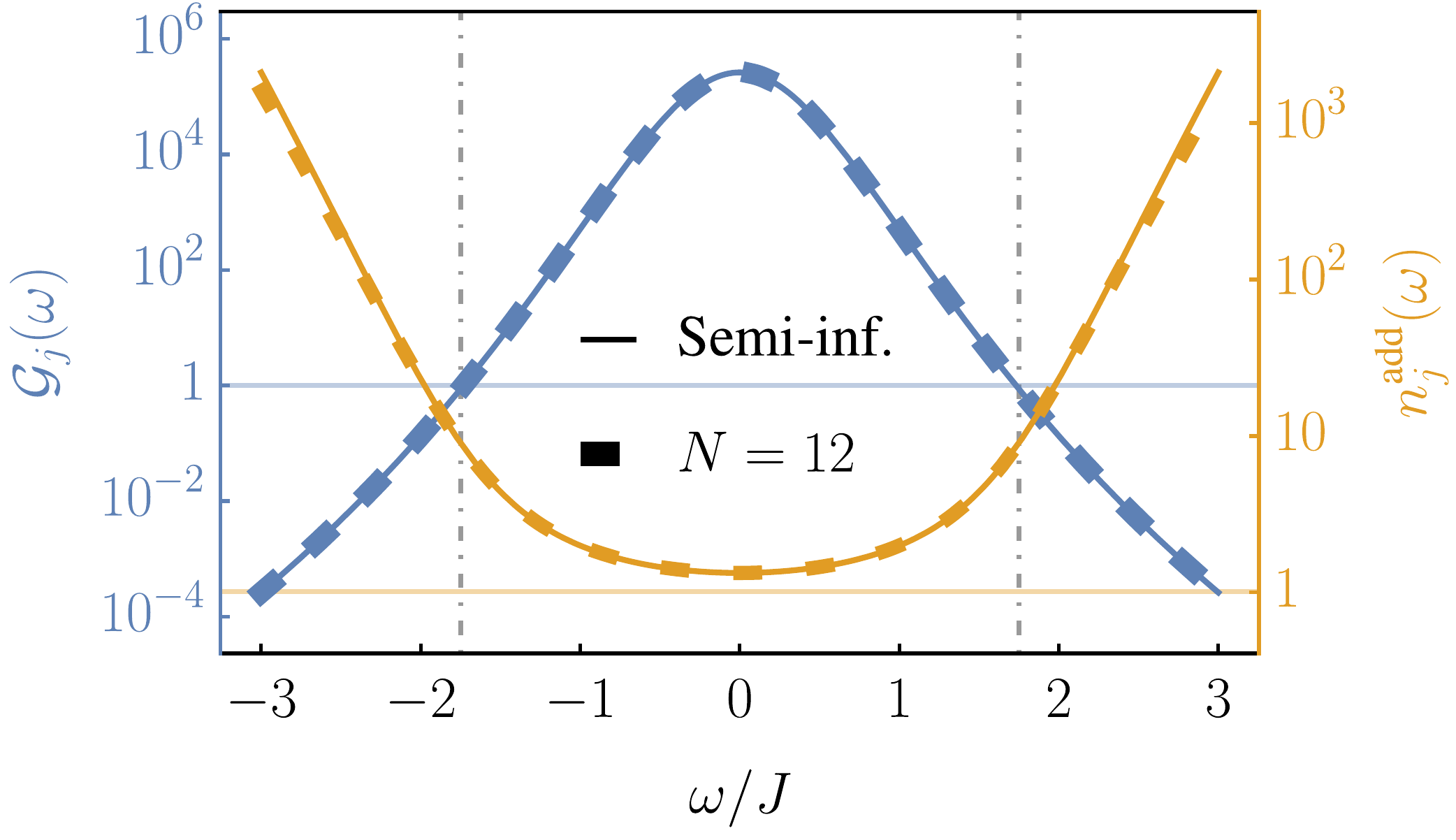}
    \caption{$\mathcal{G}_j(\omega)$~(blue) and $n_{j}^{\text{add}}\left(\omega\right)$~(yellow) at site $j=8$ for $\gamma/J=4$. The solid line corresponds to the analytical result in the semi-infinite limit, while the markers show the exact value for an array with $N=12$. The vertical dot-dashed lines indicate the phase boundaries, while the blue horizontal line, the onset for amplification. The horizontal yellow line indicates the quantum limit of noise.}
    \label{fig:Gain-Z_2}
\end{figure}

We now study the amplifier gain, which is given by Eq.~\eqref{eq:Gain}. In the semi-infinite case, we can determine its value analytically from Eq.~\eqref{eq:Projectors}.
However, we will also calculate its value numerically from the exact Green's function for a finite size array, to determine the importance of finite size effects in the amplifier properties.
Using Eq.~\eqref{eq:Projectors}, we can write for the semi-infinite case:
\begin{equation}
    \hat{G}_{j,0}(\omega)=e^{j\log \lambda_{+}}P_{+}\hat{G}_{0,0}(\omega) ,
\end{equation}
which results in the following expression for the gain at site $j$~\footnote{Remember that we have chosen $g_{s,c}/J=1$ for all our calculations, otherwise the analytical expressions change.}:
\begin{equation}
    \mathcal{G}_j(\omega)=\frac{\gamma^{2}4^{j-1}J^{2j}}{\left[\omega^{2}+\left(\frac{\gamma}{2}-J\right)^{2}\right]^{j+1}} .
\label{eq:Gain-analytical}
\end{equation}
This is shown in Fig.~\ref{fig:Gain-Z_2}~(blue), where we compare the gain at site $j=8$ from Eq.~\eqref{eq:Gain-analytical}, with the exact numerical value for a system with size $N=12$.
The analytical expression perfectly matches the result for a finite system and high gain is easily obtained for small-size arrays, confirming that the loss of stability for large systems is not important in practice.

The calculation of the noise-to-signal ratio can be carried out analogously. 
From Eq.~\eqref{eq:signal-noise-ratio}, we can particularize for this topological phase with local loss, namely:
\begin{equation}
    n_{j}^{\text{add}}\left(\omega\right)=\frac{\sum_{l=0}^{N-1}|G_{j,N+l}\left(\omega\right)|^{2}}{|G_{j,0}\left(\omega\right)|^{2}}.\label{eq:Noise-Z2}
\end{equation}
The denominator corresponds to the previous calculation of the gain, so we are left with finding the numerator, which involves the anomalous part of the Green's function that mixes positive and negative frequency modes.
It is important to note that the noise is created by all sites in the array and the not just those between the edge and site $j$. 
This means that, in the semi-infinite limit, the sum over $l$ must be extended to infinity.
In Appendix~\ref{sec:Noise} we show that this sum converges to a finite value, although, since the topological phase is directional, the propagation of signals in the opposite direction is exponentially suppressed and one could approximate the infinite sum by $l\leq j$.
In Fig.~\ref{fig:Gain-Z_2}~(yellow) we compare the analytical result in the semi-infinite limit, with the exact value for a finite array, and show the good agreement between them, only deviating at large frequencies.
This is expected, as in the trivial region directional transport is not present and the role of back scattering from the opposite boundary is relevant.

\begin{figure}
    \centering
    \includegraphics[width=1\columnwidth]{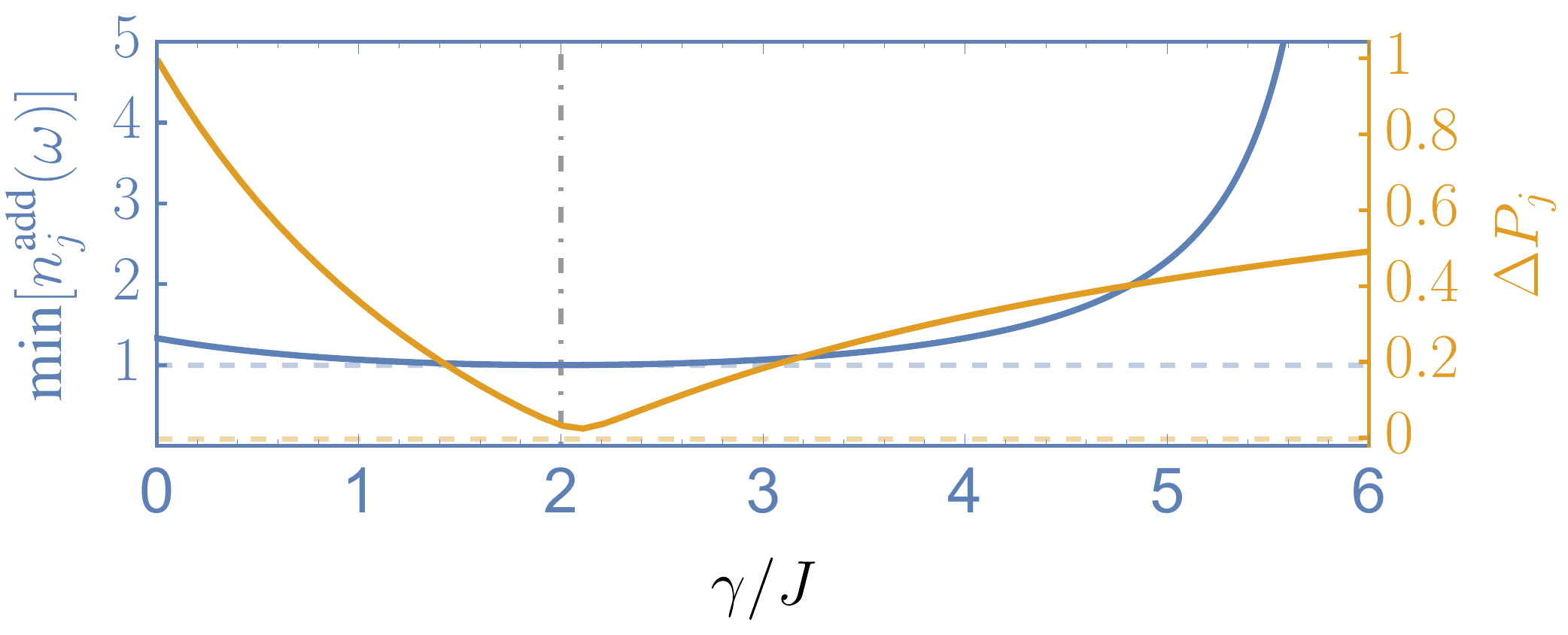}
    \caption{Blue: Minimum value of $n_{j}^{\text{add}}\left(\omega\right)$ as a function of $\gamma/J$, for a site, $j=100$, far from the edge. The horizontal dashed line indicates the quantum limit, $n_{j}^{\text{add}}\left(\omega\right)\to1$, and the vertical dashed line indicates the phase boundary of the topological phase. Yellow: Squeezing of the quadrature $\Delta P_j$ at the last site of a chain with $N=12$ and $\omega/J=0$.}
    \label{fig:Min-Noise}
\end{figure}

Importantly, one can see that the noise is close to its quantum limit, $n_{j}^{\text{add}}\left(\omega\right)\to1$~\footnote{A different convention can result in $n_{j}^{\text{add}}\left(\omega\right)\to1/2$ instead, as in ref.~\cite{QuantumLimit}}.
To quantitatively characterize this, we plot in Fig.~\ref{fig:Min-Noise}~(blue) the minimum value of the normalized added noise in the semi-infinite limit, as a function of $\gamma/J$ and for arbitrary $\omega$ within the topological phase. We find that the minimum is always at $\omega=0$ and that it reaches the quantum limit in the vicinity of the critical point, $\gamma/J\simeq2$.
Remember that there is also amplification for $\gamma/J<2$, although in this particular case the zero-energy modes are not topologically protected and the system becomes unstable.
\subsubsection{Squeezing}
The generation of squeezing in systems with parametric driving is an additional feature that can be exploited in arrays of parametric oscillators.
In contrast to the generation of squeezed states in standard parametric down conversion~\cite{Zoller-QuantumNoise}, in our case, the spatial degree of freedom of the array and its topological properties play a role.
Now we explore this phenomenon in the dissipative BdG topological phase of amplification.

\begin{figure}
    \centering
    \includegraphics[width=1\columnwidth]{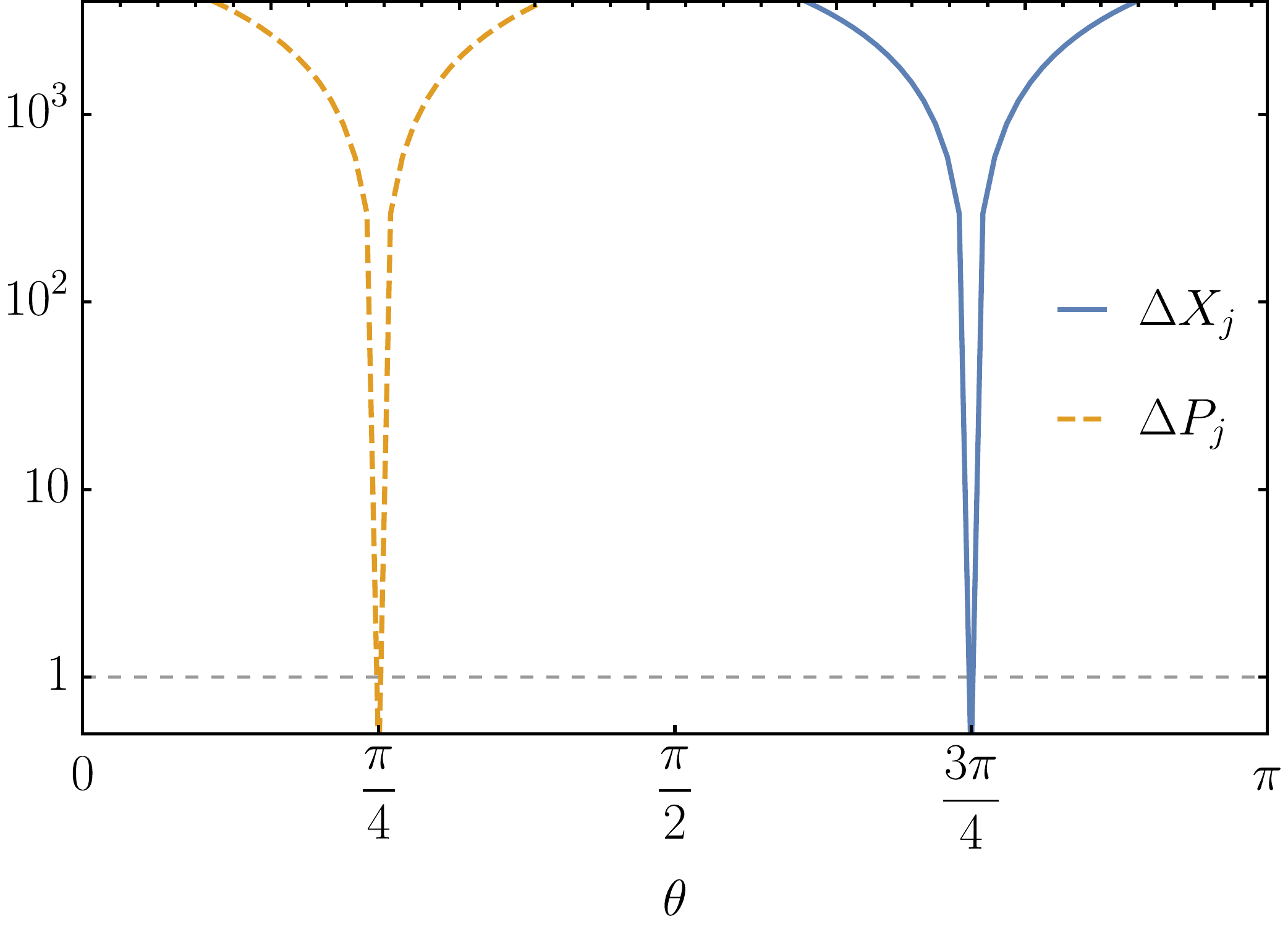}
    \caption{Variance of the quadratures as a function of $\theta$ for a finite system with $N=12$. The system is in the topological phase and the quadrature is measured at site $j=N-1$. In general, amplification increases the quadratures value several orders of magnitude, but for $\theta=\pi/4\text{ and }3\pi/4$, the quadratures are reduced below the Heisenberg limit $\Delta\mathcal{O}_j=1$~(indicated by the horizontal dashed line). Parameters: $\omega=0$, $\gamma/J=4$, $g_{s,c}/J=1$, $\phi=\pi/2$, $\Delta/J=0$ and $N=12$.}
    \label{fig:Squeezing-Theta}
\end{figure}

First, we need to find the correct angle $\theta$ for measuring the quadratures. For this, we combine Eq.~\eqref{eq:Quadratures} and the input-output relations to write the average value of the quadratures for a coherent field at site $j$:
\begin{align}
    \langle X_{j}^{\text{out}}\left(\omega_d\right)\rangle=&-i \alpha \kappa \left[  G_{j,0}\left( \omega_d \right)e^{i\theta} \right. \\
    &+ \left. G_{j+N,0}\left(\omega_d\right)e^{-i\theta} \right]  \nonumber,\\
    \langle P_{j}^{\text{out}} \left( \omega_d \right) \rangle=& \alpha \kappa \left[ G_{j,0}\left(\omega_d\right)e^{i\theta} \right. \\ 
    &\left. - G_{j+N,0} \left( \omega_d \right) e^{-i\theta} \right]. \nonumber
\end{align}
A similar result is obtained if we detect the output idler mode instead, $\omega=-\omega_d$.
We can now use the analytical expression for the Green's function in the topological phase, Eq.~\eqref{eq:Projectors}, to factor out the $\theta$-dependence and see that $\langle X_{j}^{\text{out}}\left(\omega_d\right)\rangle\propto \cos(\theta)+\sin(\theta)$ and $\langle P_{j}^{\text{out}}\left(\omega_d\right)\rangle\propto \cos(\theta)-\sin(\theta)$. This means that for $\theta=\pi/4$ the $\langle X_{j}^{\text{out}}\left(\omega_d\right)\rangle$ quadrature is maximized, while $\langle P_{j}^{\text{out}}\left(\omega_d\right)\rangle$ vanishes.
Notice that this is a direct consequence of the structure of the projectors in Eq.~\eqref{eq:Projectors}.
%%%%
%For this, we compare with the standard result in parametric down conversion, which is a limit that should be recovered in the absence of hopping~\cite{Zoller-QuantumNoise}.
%There, squeezing around $\omega=0$ is generated for quadratures with $\theta=\pi/4$, where the anomalous terms of the Hamiltonian take the canonical form used in parametric-down conversion $H_{\text{p-d}}\propto i g_{s}(a^{2} - a^{\dagger 2})$.
To confirm that the direction in which the quadrature $\langle P_{j}^{\text{out}}\left(\omega_{d}\right)\rangle$ is squeezed coincides with the direction where its average vanishes, we show in Fig.~\ref{fig:Squeezing-Theta} that the variance in any of the quadratures can be minimized  at $\theta=\pi/4\mod (\pi/2)$, to values below the Heisenberg limit. Hence, from now on we fix the direction of the quadratures to $\theta=\pi/4$, where the $P_j$ quadrature can be squeezed.

\begin{figure}
    \centering
    \includegraphics[width=1\columnwidth]{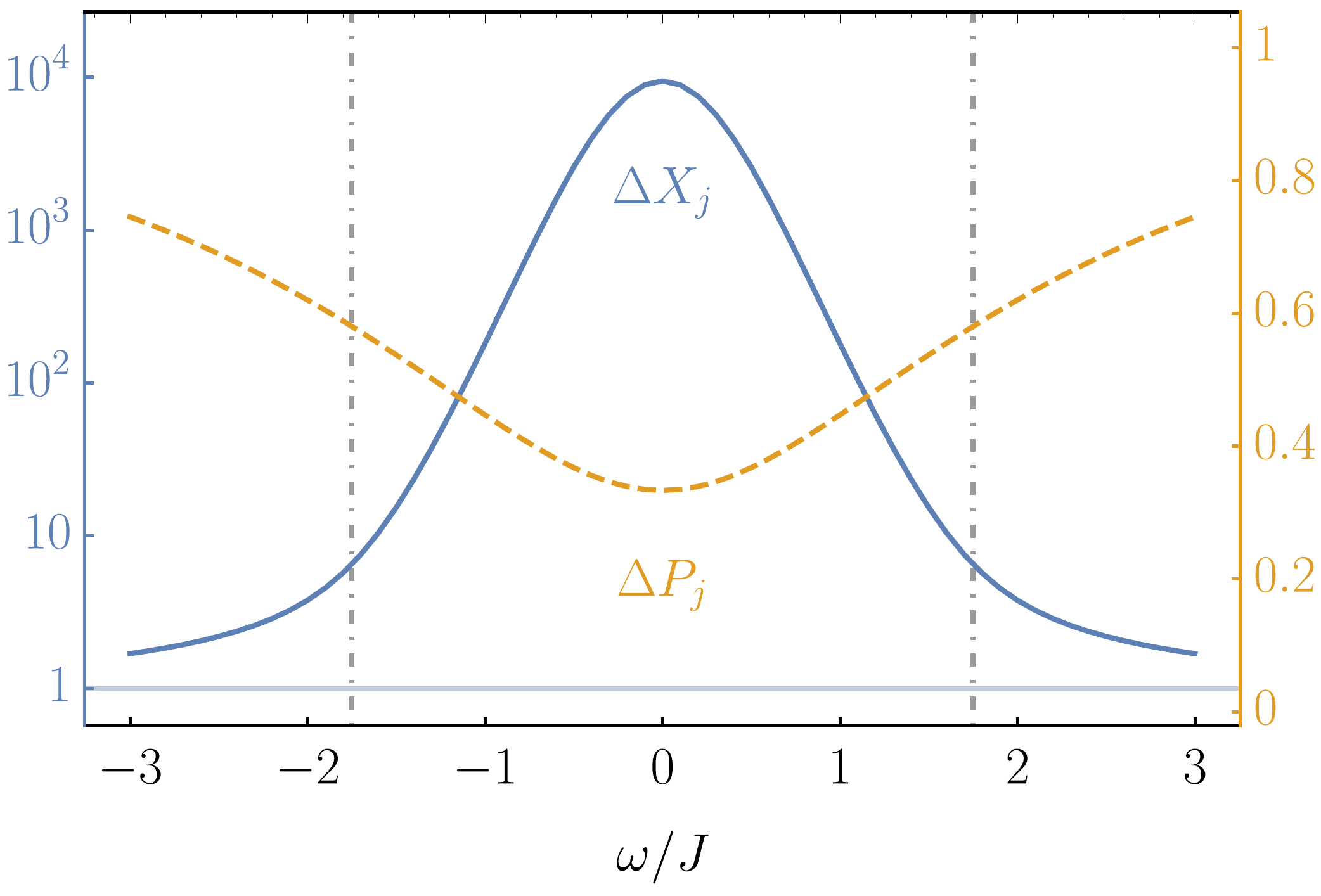}
    \caption{Variance of the quadratures at the edge of the array, $j=N-1$, as a function of $\omega/J$ and for the BdG topological phase. The $X$ quadrature is amplified while the $P$ is squeezed. Parameters: $\gamma/J=4$, $g_{s,c}/J=1$, $\phi=\pi/2$, $\Delta/J=0$ and $N=12$.}
    \label{fig:Squeezing-omega}
\end{figure}

\begin{figure}
    \centering
    \includegraphics[width=1\columnwidth]{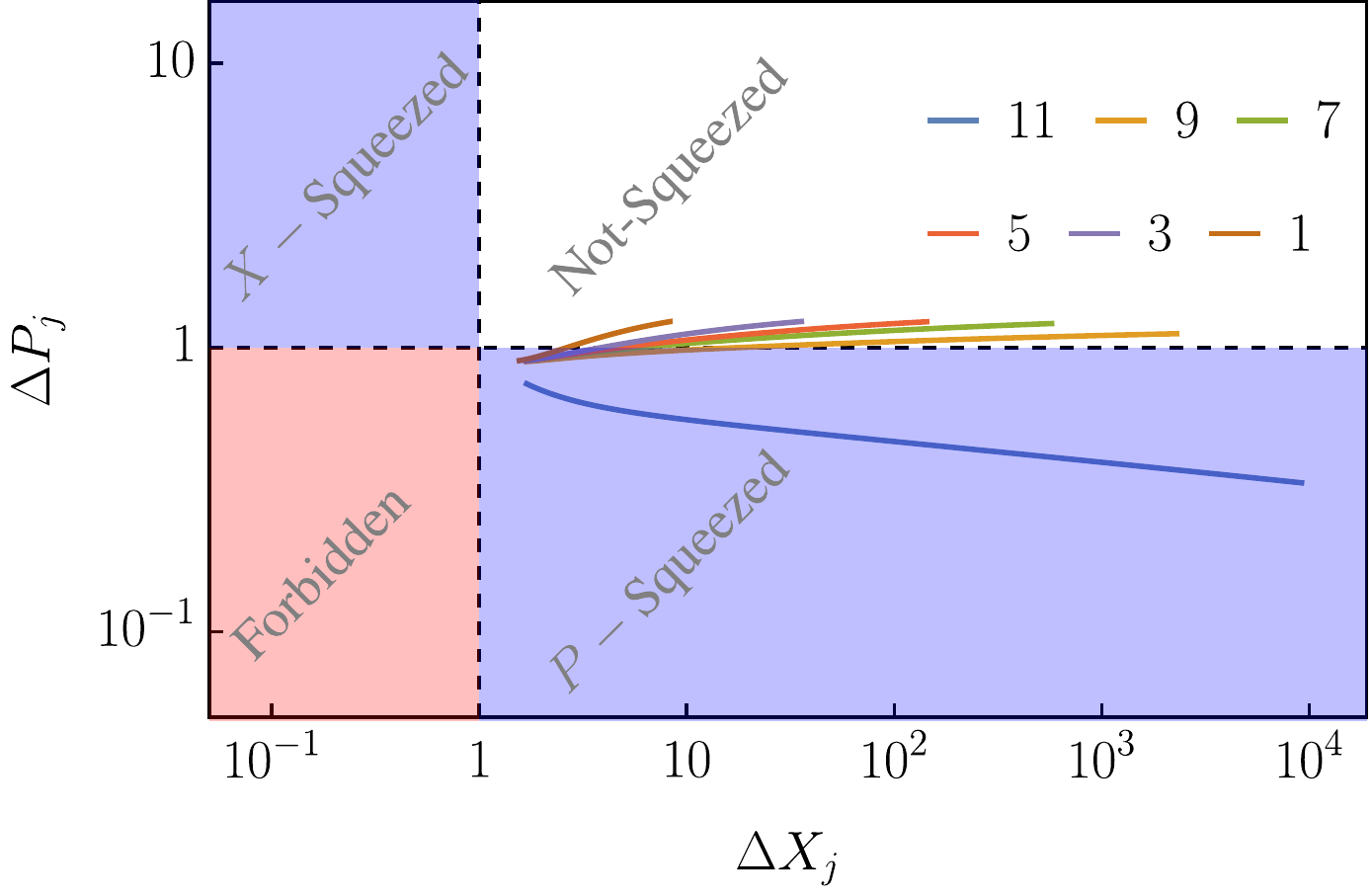}
    \caption{Squeezing diagram which shows the trajectories, as a function $\omega$, of the vector $[\Delta X_j(\omega),\Delta P_j(\omega)]$ for different sites $j=1,3,\ldots,11$. The last site trajectory is within the region of momentum squeezed states, as expected from Fig.~\ref{fig:Squeezing-omega}. Parameters: $\gamma/J=4$, $g_{s,c}/J=1$, $\phi=\pi/2$, $\Delta/J=0$ and $N=12$.}
    \label{fig:Squeezing-diagram}
\end{figure}

In Fig.~\ref{fig:Squeezing-omega} we show how the variance of each quadrature depends on frequency. 
One can see that one of the quadratures is amplified, as expected for the topological amplification phase, while the other is squeezed for a wide range of frequencies. The plot is shown for the last site of the array, $j=N-1$, but if we consider intermediate sites, where amplification is reduced, we find that the squeezing also decreases.
In contrast, increasing the size of the array boosts amplification for the corresponding quadrature but it does not significantly affect the squeezing in the other one.
All these properties are encapsulated in the diagram of Fig.~\ref{fig:Squeezing-diagram}. There, each trajectory corresponds to the one followed by the vector $[\Delta X_j(\omega),\Delta P_j(\omega)]$, as a function of $\omega$. 
The diagram is divided in four regions: the red one is forbidden by the Heisenberg limit $\Delta X_j \Delta P_j\leq 1$, the blue ones correspond to one of the quadratures being squeezed and the white one corresponds to states which are not squeezed along any particular direction.
We find squeezing for the $P_j$ quadrature in the dissipative BdG topological phase. For the last site, $j=N-1$, is present for all frequencies. However, one can see important differences between sites, as squeezing is rapidly reduced as $j$ approaches $j=0$. In contrast, the $X_j$ quadrature is always amplified several orders of magnitude.\\
Finally, we plot in Fig.~\ref{fig:Min-Noise}~(yellow), the maximum squeezing that can be obtained in this topological phase. It follows a similar behavior as the noise, approaching the ideal limit, $\Delta P_j \to 0$, in the proximity of the "critical point", $\gamma/J=2$, where the amplification of $\Delta X_j$ also grows.
\subsection{Double Hatano-Nelson phase ($P\neq 0$)}
As mentioned in Sec.~\ref{sec:Topology}, the presence of collective pump makes possible to reach $W_1(\omega)=2$ and produces a larger number of topologically protected zero-energy modes.
We now study the amplifier properties in this phase.
\subsubsection{Coherence length}
The inverse coherence length in this topological phase can be studied analogously.
However, due to $P\neq0$, the non-linear matrix equation for the surface Green's function is now more complicated and makes more difficult to find compact analytical expressions.
For this reason and because the region of topological stability in Fig.~\ref{fig:Stability2} is small and depends on size, we will restrict our analysis to a fully numerical one for the case $P/J=0.75$ and $\gamma/J=4$, which is indicated by the red dot in Fig.~\ref{fig:Stability2}. 
Nevertheless, it is important to mention that as in the previous case, the comparison between the finite system and the semi-infinite limit always shows good agreement.
%This is also adequate in general for $t_d/J\neq0$, due to the dominance of unstable regions and their strong size dependence.
\begin{figure}
    \centering
    \includegraphics[width=1\columnwidth]{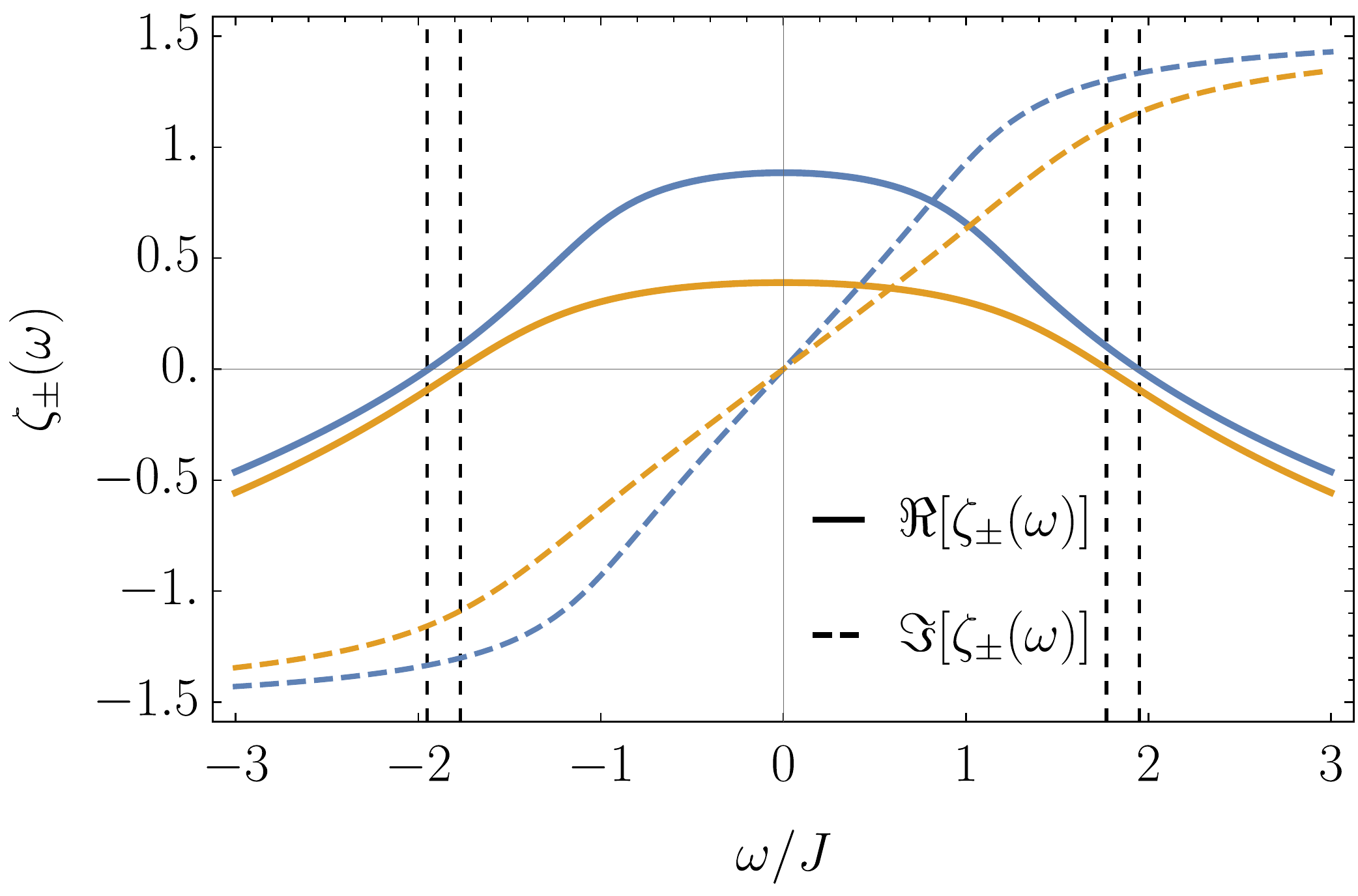}
    \caption{$\zeta_{\pm}(\omega)$ as a function of $\omega/J$ for $g_{s,c}/J=0.1$, $\phi=\pi/2$, $\Delta/J=0$, $P/J=0.75$ and $\gamma/J=4$. Regions of amplification correspond to $\Re [\zeta_{\pm}(\omega)]> 0$. Vertical dashed lines indicate changes in the topological invariant from the phase diagram in Fig.~\ref{fig:PhaseDiagramZ}.}
    \label{fig:CoherenceLength2}
\end{figure}

The calculation of the inverse coherence length shows that this topological phase is characterized by amplification in both subspaces.
This is shown in Fig.~\ref{fig:CoherenceLength2}, where the condition $\Re [\zeta_{\pm}(\omega)]> 0$ determines the regions of topological amplification.
Notice that the critical points in the phase diagram of Fig.~\ref{fig:PhaseDiagramZ} can be identified with the frequency values were amplification starts for each $\zeta_{\pm}(\omega)$.
It is also interesting to notice that, if we continuously reduce the parametric terms, both inverse coherence lengths approach each other without modifying the winding number, until they are indistinguishable.
As for $g_{s,c}/J=0$ the positive and negative frequency modes decouple, this confirms that this topological phase can be identified with two weakly coupled copies of the Hatano-Nelson model.
Next, we will show that this has consequences in the gain and noise produced by the amplifier, but the squeezing properties will be the ones primarily affected.
\subsubsection{Amplifier gain and noise}
\begin{figure}
    \centering
    \includegraphics[width=1\columnwidth]{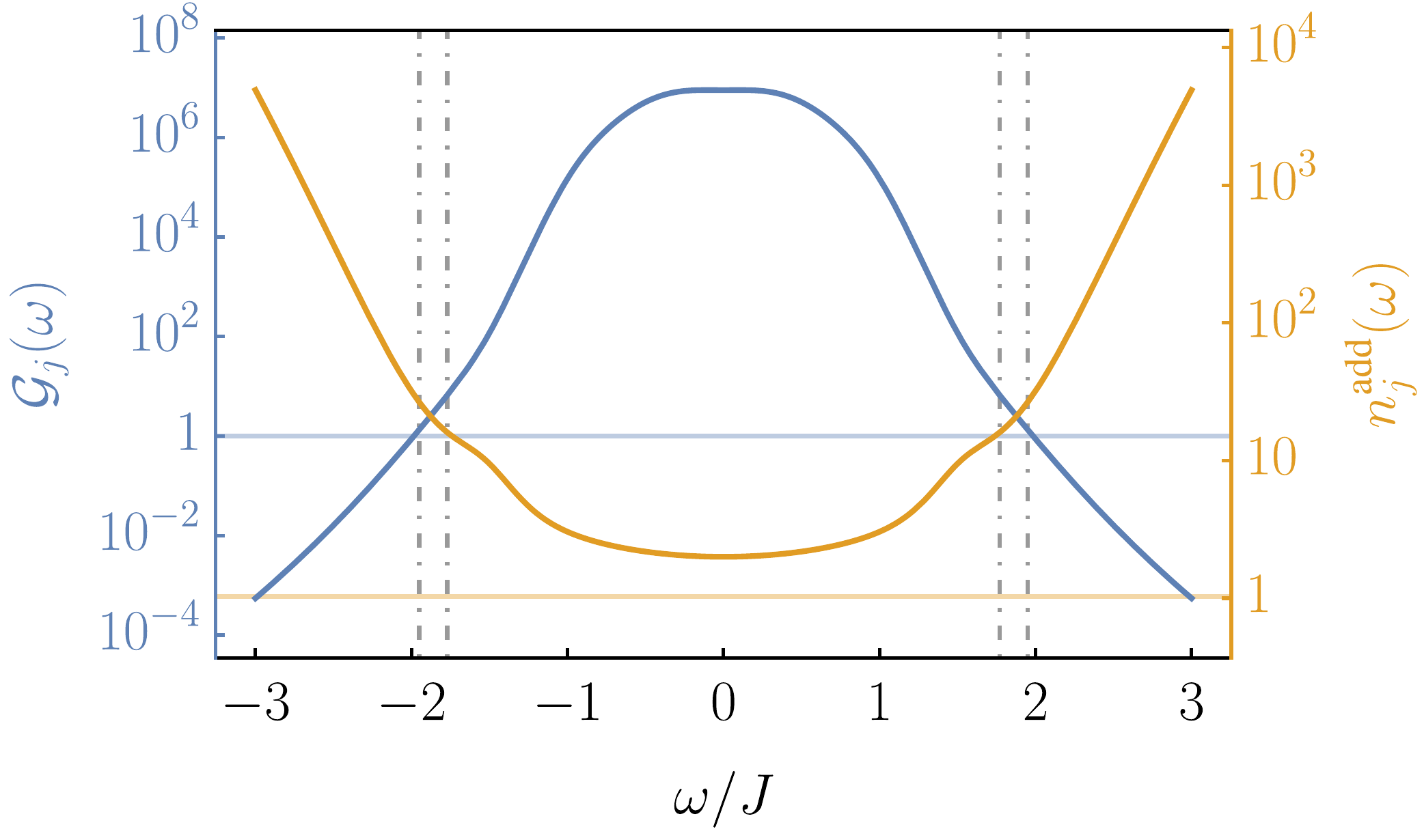}
    \includegraphics[width=1\columnwidth]{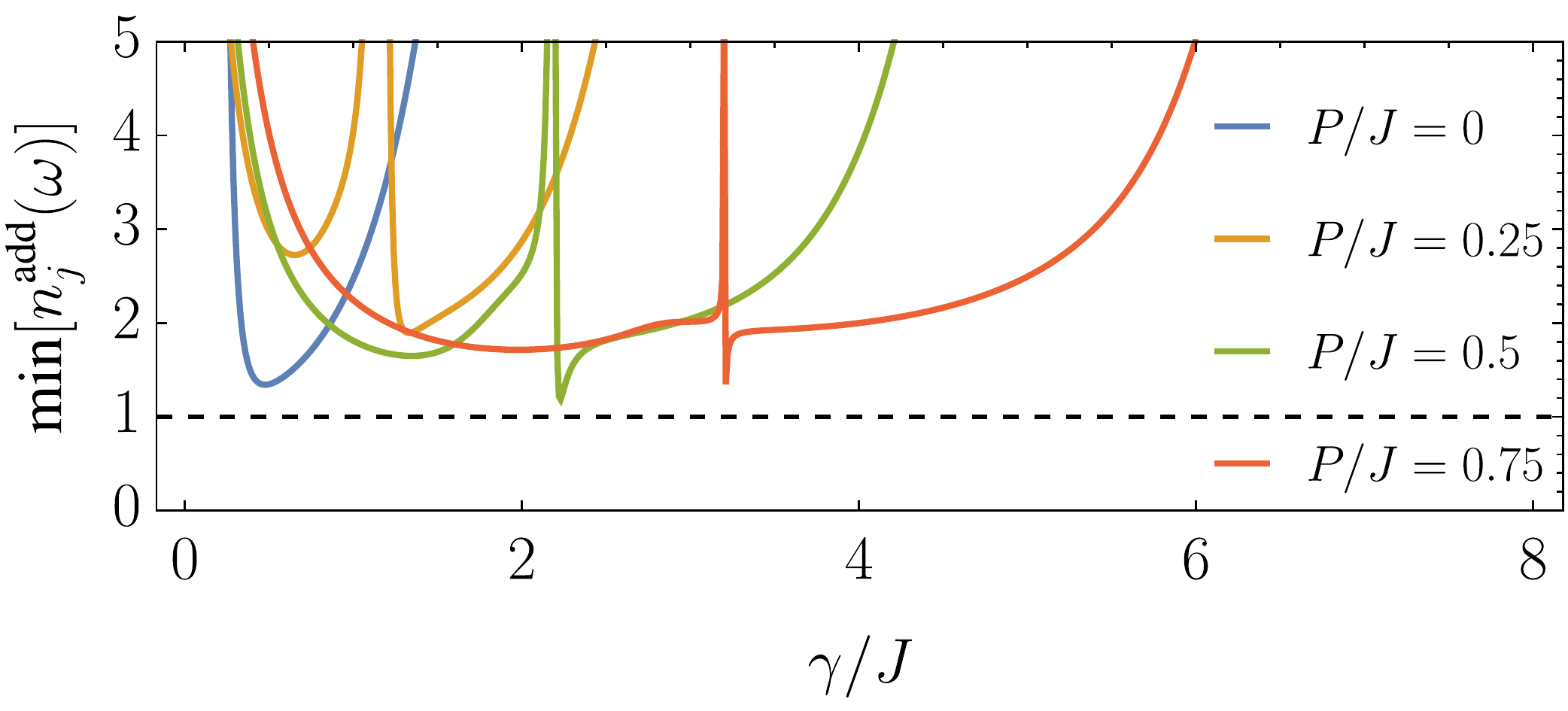}
    \caption{Top: Gain and noise at site $j=8$ for the topological phase with $W_1=2$, as a function of $\omega$ and for an array with $N=12$ sites. The vertical dot-dashed lines indicate the regions where the topological invariant changes. The horizontal lines indicate the onset of amplification (blue) and the quantum limit of noise (yellow).
    Bottom: Noise minimum value vs $\gamma/J$ for different values of $P/J$. The dashed line indicates the quantum limit for the noise.}
    \label{fig:Gain-Z}
\end{figure}

We plot in Fig.~\ref{fig:Gain-Z}~(top, blue), the gain at site $j=8$ for a finite array with $N=12$ sites.
The gain is greater than in the topological phase with $W_{1}(\omega)=1$, which can be attributed to the presence of amplification in both subspaces.
As expected from our formalism, the region of amplification, marked by vertical dot-dashed lines, perfectly matches the boundaries of the topological phase diagram in Fig.~\ref{fig:PhaseDiagramZ}.
Regarding the noise-to-signal ratio in this topological phase, the expression is now slightly more complex than in Eq.~\eqref{eq:Noise-Z2} due to the additional collective pump term:
\begin{align}
    n^\text{add}_j(\omega)=&\frac{\sum_{l=0}^{N-1}|G_{j,N+l}\left(\omega\right)|^{2}
    }{|G_{j,0}\left(\omega\right)|^{2}}\nonumber\\
    &+\frac{\sum_{l,l^{\prime}=0}^{N-1} P_{l^{\prime},l}G_{j,l}^{*}\left(\omega\right)G_{j,l^{\prime}}\left(\omega\right)}{\gamma |G_{j,0}\left(\omega\right)|^{2}}.\label{eq:Noise-Z}
\end{align}
However, its behavior in Fig.~\ref{fig:Gain-Z}~(top, yellow) also shows that noise is strongly reduced in the topological phase.
The main difference with the dissipative BdG topological phase is that in this case, the noise does not reach the quantum limit (cf Fig.~\ref{fig:Gain-Z_2}).
%In addition, for our choice a parameters to obtain a stable region, there is a small range of frequencies between critical points where $W_1(\omega)=1$.
To explore in more detail the possibility to reach the quantum limit, we plot in Fig.~\ref{fig:Gain-Z}~(bottom), the minimum value of $n^\text{add}_j(\omega)$, as a function of the dissipative parameters $P/J$ and $\gamma/J$.
It shows that the minimum shifts as $P/J$ changes, but the quantum limit is not reached.
If we compare with the stability phase diagram in Fig.~\ref{fig:Stability2}, one can see that the noise minimum always coincides with the transition from the unstable to the stable topological region, as it also happened in the dissipative BdG topological phase (Fig.~\ref{fig:Min-Noise}), however, in this case the minima coincide with a divergence. This is a consequence of the finite size of the system, and we have checked that the divergence softens as the size of the system increases and the noise is detected further away from $j=0$. Concretely, we have seen that increasing the size of the system for $P/J=0.75$, saturates $\text{min}[n_N^{\text{add}}(\omega)]\to 1.95$, as $N\to\infty$.
\subsubsection{Squeezing}
\begin{figure}
    \centering
    \includegraphics[width=1\columnwidth]{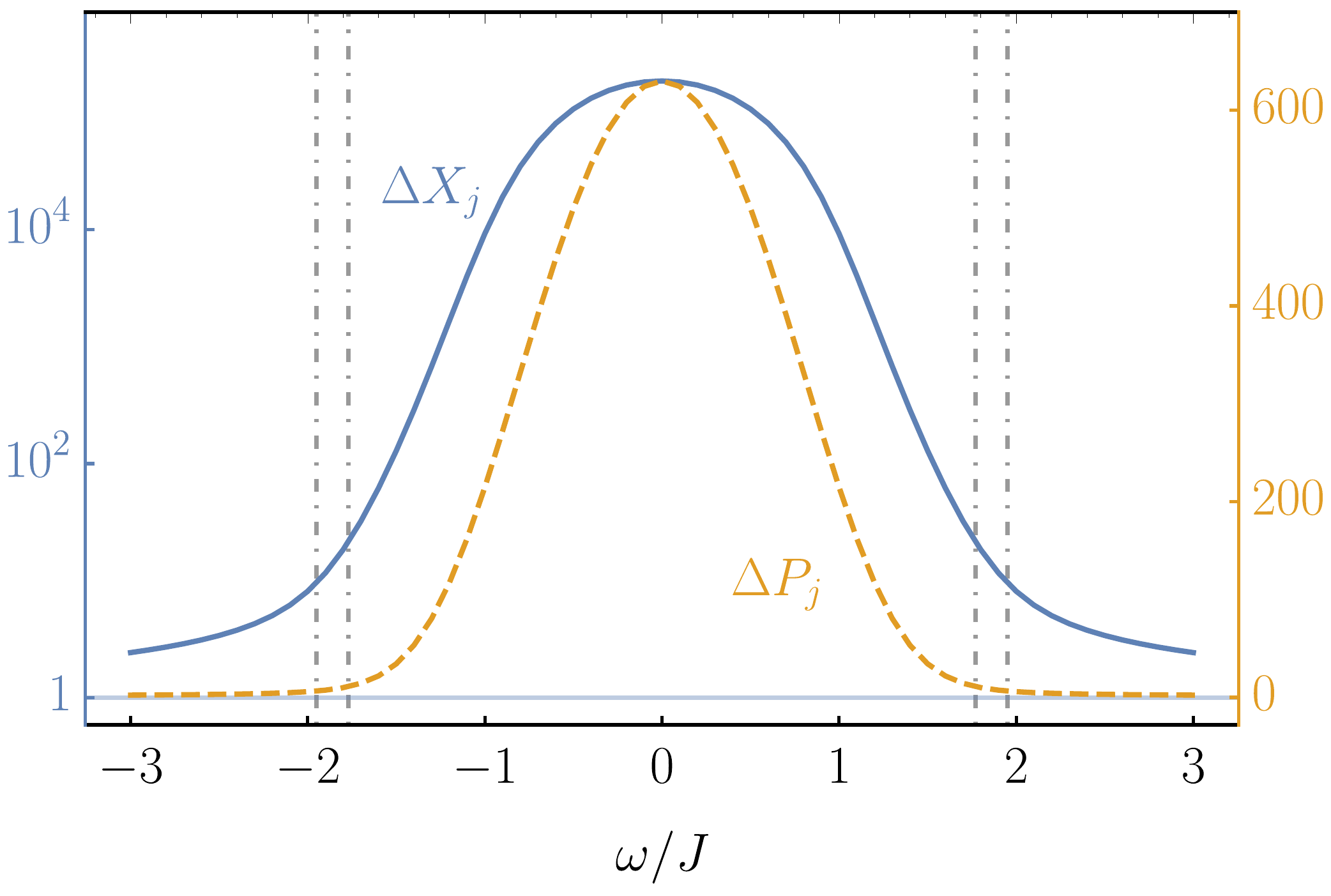}
    \includegraphics[width=1\columnwidth]{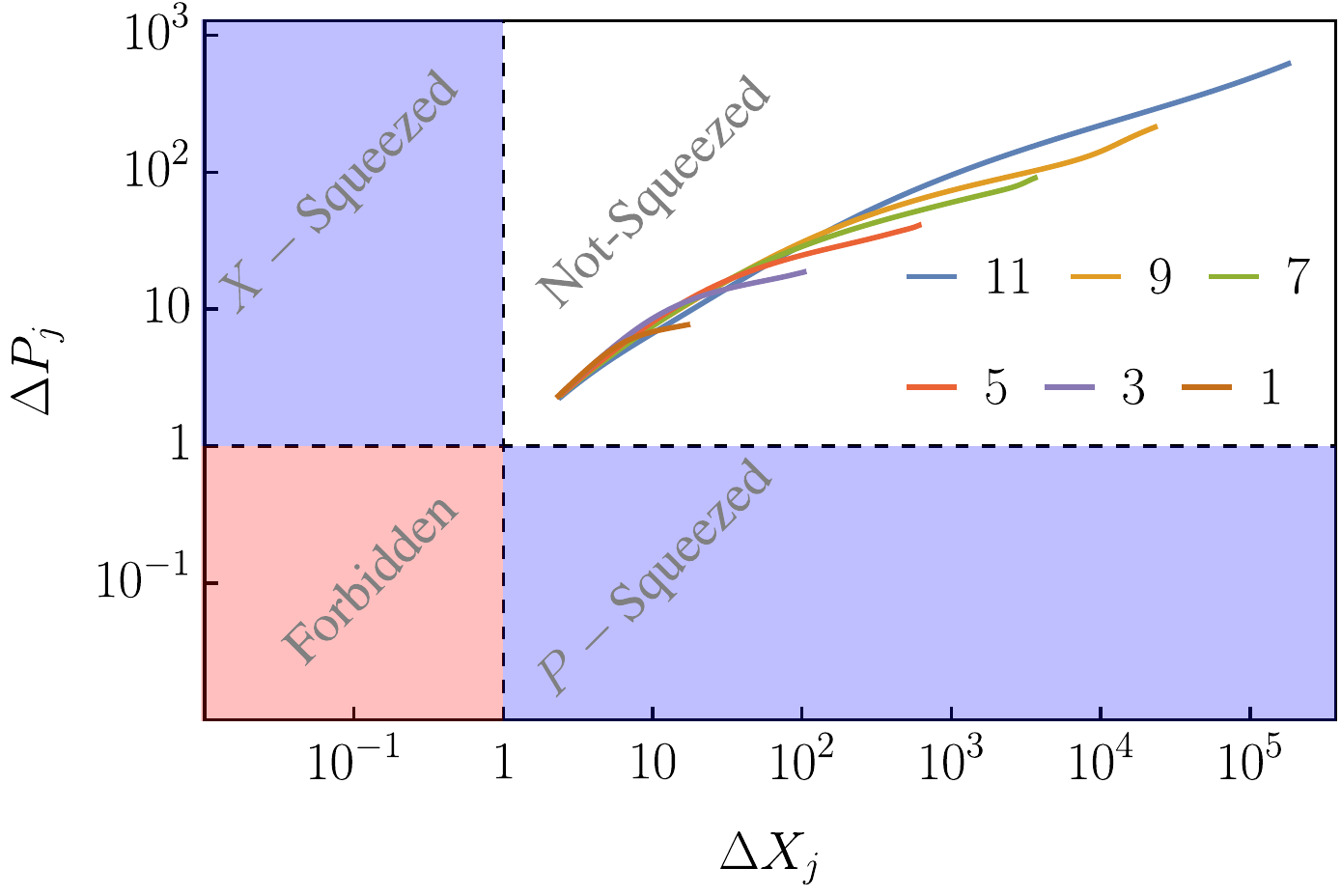}
    \caption{Top: Variance of the quadratures as a function of $\omega$ for the topological phase with $W_1=2$. Parameters: $g_{s,c}/J=0.1$, $\phi=\pi/2$, $\Delta/J=0$, $P/J=0.75$ and $\gamma/J=4$.
     Bottom: Trajectories of the vector $[\Delta X_j(\omega),\Delta P_j(\omega)]$ for different sites $j=1,3,\ldots,11$ and the same parameters as the top plot.}
    \label{fig:Squeezing-Z}
\end{figure}

Squeezing is greatly affected in this topological phase with collective pump.
This is shown in Fig.~\ref{fig:Squeezing-Z}~(top), where one can see that both quadratures are simultaneously amplified. The different ratio of amplification between quadratures is produced by the unequal correlation length in each subspace (see Fig.~\ref{fig:CoherenceLength2}).
Importantly, we have checked that the lack of squeezing in this topological phase is not due to a rotation of the quadratures, as we have checked that the minimum of each variance is still found for $\theta=\pi/4$.
This result is to be expected due to the similarities between this topological phase and the one in the Hatano-Nelson model, where squeezing cannot be created. 
In this case, there is a small contribution from $g_{s,c}/J=0.1$, but this is not enough to produce relevant correlations between positive and negative frequency modes, required to generate squeezed states.
 Moreover, Fig.~\ref{fig:Squeezing-Z}~(bottom) shows that squeezing is not produced in any of the sites of the array. All trajectories of the vector are away from the region of squeezed states. Their different length along each direction is a direct consequence of the unequal correlation length in each subspace, while their change in length between sites shows that amplification happens exponentially with the number of sites.
 These results confirm that the topological phase with collective pump can also be used as a good amplifier, due to its large gain, stability, great resilience to disorder and small noise-to-signal ratio, although it lacks the possibility to generate squeezing.
\section{Experimental implementations\label{sec:Experiments}}
The dissipative topological phases of amplification can be implemented in a variety of quantum optical and nano-mechanical setups, including superconducting qubits \cite{Houck2012,Ripoll2022}, trapped ions \cite{Schneider2012,Blatt2012}, and opto-mechanical systems \cite{Marquardt2013}. 
The main requirements are the ability to induce or control squeezing terms in the Hamiltonian together with  complex phases in the photon hopping energies. We focus below in two implementations. The first one takes advantage of the presence of controllable Kerr non-linearities in superconducting circuits. The second scheme relies on multi-tone drivings to control coupling terms that are naturally present in systems of coupled microwave or nano-mechanical resonators, as well as in trapped ion systems. 

\subsection{Implementation with Kerr non-linearities}
Parametric terms such as those that produce squeezing in Eq.~\eqref{eq:Hamiltonian}, can arise from three- or four-wave mixing processes, which occur as a result of non-linear effects. The latter are naturally present in non-linear materials, however, they can also be engineered in superconducting circuits by using Josephson junctions.

Let us focus on the case of four-wave mixing. Here, squeezing terms can be produced by means of the Kerr non-linear effect, which leads to two types of terms:
\begin{eqnarray}
    H_s &=& \sum_j K_{s} (a_j + a_j^\dagger)^4 , \nonumber \\
    H_c &=& \sum_j K_{c} (a_j + a_j^\dagger)^2(a_{j+1} + a_{j+1}^\dagger)^2.
\end{eqnarray}
The parametric driving of a chain of coupled Kerr oscillators at frequencies $2 \omega_p$ will bring terms like $a^\dagger_j a^\dagger_{j}$,  $a^\dagger_j a^\dagger_{j+1}$, into resonance.
This is the usual method followed with superconducting circuits for squeezing generation, for example, in Refs.~\cite{roy_introduction_2016,eichler_controlling_2014}. 

The normal tunneling terms, on the other hand, appear quite naturally in systems of coupled microwave superconducting cavities. The phase $\phi$ can be controlled, for example,  with Floquet engineering techniques. 
In Ref.~\cite{Ramos2022} we present details on how to implement our dissipative BdG topological phase by using Kerr non-linearities in a realistic superconducting setup, without the need of Floquet engineering or magnetic fields to break time-reversal symmetry.
Finally, the squeezing terms that are required for our scheme, can also be implemented with three-wave mixing terms, as explained in Ref.~\cite{McDonald2018}.
\subsection{Implementation with periodic driving of the local frequencies}
\subsubsection{Periodic driving of the local frequencies}
Our topological array of parametric oscillators could be implemented with a Floquet-engineering scheme that does not require the presence of non-linearities. 
This approach is appropriate for systems of coupled microwave resonators, as well as for nanomechanical or trapped ion setups, in which vibrational excitations play the role of the photons in the amplifier.  
Our proposed method relies on the periodic driving of the local resonator frequencies, an idea successfully demonstrated in the vibrational case with trapped ions in Ref.~\cite{KieferPRL2019}.

Consider a chain of coupled resonators described by the following time-dependent Hamiltonian,
\begin{equation}
H(\tau) = H_0  + V + H_d(\tau).
\end{equation}
Here, $H_0$ describes the local frequencies, which we assume to be equal,
\begin{equation}
H_0 = \sum_j \omega_{r} a^\dagger_j a_j,
\label{H.local}
\end{equation}
where $a_j^\dagger$ and $a_j$ can be photonic or phononic operators. 
In the case of trapped ions, $H_0$ describes the quantized oscillations of ions around their equilibrium positions. 
$V$ is the coupling between resonators and takes the form of a position- or field-dependent interaction, which we write in terms of dimensionless position operators $X_j = (a_j + a_j^\dagger)/\sqrt{2}$:
\begin{eqnarray}
V &=& 
\sum_{j} J_c \left(X_j - X_{j+1} \right)^2 \nonumber \\
&=&
2 \sum_j J_c X_j^2 - 2 \sum_j J_c X_j X_{j+1} \nonumber \\
&=& 
\sum_j J_c \left(a_j + a_j^\dagger \right)^2
\nonumber \\
&-&
\sum_{j} J_c \left(a_j +a_j^\dagger \right) \left(  a_{j+1} + a_{j+1}^\dagger \right), 
\label{dipolar}
\end{eqnarray}
where we have assumed that only nearest-neighbours are coupled. This is the natural situation for systems of microwave or nanomechanical resonators. In the case of trapped ions, vibrational couplings decay fast with distance, so that the nearest-neighbour model is still a good approximation, although the formalism presented here could be easily adapted to account for long-range terms.

The time-dependent term consists of a periodic driving of the local energies with driving frequency $2 \omega_r$, 
\begin{eqnarray}
   H_d(\tau) &=& \sum_{j} 2 \eta \omega_r \cos( 2 \omega_r \tau + \phi_j)
   a_j^\dagger a_j,
\end{eqnarray}
with $\eta$ the driving strength relative to the frequency, $\omega_r$. The modulation frequency has been chosen as twice the local frequency to put into resonance the squeezing terms. Finally, we need to have a position-dependent driving phase:
\begin{equation}
    \phi_j = j \Delta \phi \ ,
\end{equation}
for reasons that will become clear below.

To see  how the scheme works, we define the evolution operator in the interaction picture with respect to $H_0 + H_d(\tau)$,
\begin{equation}
    U(t) = \exp\left( - i \int_0^t \left(H_0 + H_d(\tau) \right) d\tau \right),
\end{equation}    
in which bosonic operators evolve like,
\begin{eqnarray}
    \bar{a}_{j}^\dagger(t) =  U^\dagger(t) a_j^\dagger U(t) 
                           = a^\dagger_j e^{i \omega_r t}  
                            e^{i \eta \sin(2 \omega_r t + \phi_j)}.
\end{eqnarray}
and have ignored a phase factor, $e^{i \eta \sin(\phi_j)}$, that can be trivially gauged away. 

We proceed by writing coupling terms in the interaction picture defined above. We assume that $\omega_r \gg J_{c}$, so that we neglect any nonresonant terms in a rotating wave approximation.
However, we do not need to work in a perturbative regime of small $\eta$. We will make use of the following identity: 
\begin{equation}
    e^{i\eta \sin(2\omega_r t + \phi_j)} = \sum_{n = 0, \pm 1, ...} {\cal J}_n(\eta) e^{i n  2 \omega_r t + i n \phi_j},
\end{equation}
where ${\cal J}_n(\eta)$ are the Bessel functions of the first kind. Let us start by writing the squeezing terms in the interaction picture:
\begin{eqnarray}
        \bar{a}_j^\dagger \bar{a}^\dagger_{j}  
        &=& a_j^\dagger a_j^\dagger e^{2 i \omega_r t} \sum_{n} {\cal J}_n(2 \eta) 
        e^{i n 2 \omega_r t + i n \phi_j}  \nonumber \\
        &\approx&
        a_j^\dagger a_j^\dagger  {\cal J}_{-1}(2 \eta) e^{- i \phi_j} ,
\end{eqnarray}
where the approximation implies neglecting terms evolving with frequencies that are multiples of $\omega_r$. We follow the same procedure with two-site squeezing terms
\begin{eqnarray}
        \bar{a}_j^\dagger \bar{a}^\dagger_{j+1}  
        &\approx&
        a_j^\dagger a_{j+1}^\dagger \sum_n {\cal J}_{-n-1}(\eta) {\cal J}_n(\eta) 
        e^{i n \phi_j} e^{-i (n+1) \phi_{j+1}} 
        \nonumber \\
        &=&         
        a_j^\dagger a_{j+1}^\dagger 
        F(\eta, \Delta \phi) e^{- i \phi_j/2} e^{- i \phi_{j+1}/2}\ ,
\end{eqnarray}
where we have defined the function: 
\begin{equation}
F(\eta,\Delta \phi) 
= \sum_n {\cal J}_{-n-1}(\eta) {\cal J}_n(\eta) e^{- i (n + 1/2) \Delta \phi} ,
\end{equation}
which can be easily checked to be real-valued. Finally, the normal boson hopping terms are also renormalized:
\begin{eqnarray}
        \bar{a}_j^\dagger \bar{a}_{j+1}  
        &\approx&
        a_j^\dagger a_{j+1} \sum_n \left[ {\cal J}_{n}\left(\eta\right)\right]^2 e^{-i n \Delta \phi} .
\end{eqnarray}
Again, it can be easily shown that the obtained factor is real-valued.

The periodic drivings, thus, leads to a dressed system where the coupling Hamiltonian in Eq.~\eqref{dipolar} is transformed into the Hamiltonian part of our array of parametric oscillators.
Remarkably, we will get squeezing Hamiltonian terms that have a real value, up to a complex site-dependent phase. 
To remove them, we can perform a gauge transformation. 
\begin{equation}
a_j \to a_j e^{i \phi_j / 2} ,   
\end{equation}
so that the phases disappear from the squeezing terms. However in the new gauge, the phase difference appears in the normal boson hopping terms. 
Thus, we finally get the desired Hamiltonian with constants:
\begin{eqnarray}
    J &=&   - J_c  \sum_n \left[ {\cal J}_{n}(\eta) \right]^2 e^{-i n \Delta \phi}  ,
    \nonumber \\ 
    \phi &=& \Delta \phi/2 ,
    \nonumber \\
    g_{s}  &=&  J_c {\cal J}_{-1}(2 \eta)  \nonumber \\
    g_{c} &=&   - J_c F(\eta, \Delta \phi) .
\end{eqnarray}
We emphasize that the periodic drivings required for our scheme are feasible, for example with trapped ion technologies, where all the elements required for this proposal have been demonstrated, in particular the control of hopping terms to add a synthetic gauge field was theoretically proposed in Refs.~\cite{Ions1,Bermudez_2012} and recently demonstrated in an experiment~\cite{KieferPRL2019}. 
Other possibilities, such as non-perturbative driving regimes or exploiting anharmonic mechanical terms in much the same way as Kerr nonlinearities, could be explored.
Finally, dissipative terms in Eq.~\eqref{Ld} can be implemented by means of laser cooling in the case of trapped ions, or they would provided by natural decay in the case of nano-mechanical or microwave resonators. 

\subsubsection{Periodic driving of the couplings}

An alternative mechanism that can allow us to implement the topological amplification phase relies on the periodic driving of the coupling terms, rather than the local frequencies. 
This scheme has been used, for example, in the case of microwave superconducting resonators in Ref.~\cite{roushan17}. In trapped ion setups, this method would be more challenging, since it would require displacing ions to modify the Coulomb coupling. However, parametric resonance between different vibrational modes has already been demonstrated in Ref.~\cite{Haffner14pra}. 

The scheme would work by periodically driving the coupling terms. Our Hamiltonian takes the form,
\begin{eqnarray}
 H = H_0 + V(t),   
\end{eqnarray}
In order to have full control on the final Hamiltonian, it is advantageous to have a gradient of the resonator frequencies, so that, now,
\begin{equation}
    H_0 = \sum_j \omega_j a_j^\dagger a_j,
\end{equation}
with $\omega_j = \omega_0 + j \Delta \omega$.
$V(t)$ has the same form as in Eq.~\eqref{dipolar}, but with a time- and site-dependent coupling,
\begin{eqnarray}
V(t) &=&
\sum_j J_{c,j}(t) \left(a_j + a_j^\dagger \right)^2
\nonumber \\
&+&
\sum_{j} J_{c,j}(t) \left(a_j +a_j^\dagger \right) \left(  a_{j+1} + a_{j+1}^\dagger \right).    
\label{dipolar2}
\end{eqnarray}
We consider the following multi-tone driving for the couplings:
\begin{eqnarray}
    J_{c,j}(t) &=&
    A_0 +
    A_{1} \cos(2 \omega_j t) +  
    A_{2} \cos[(\omega_j + \omega_{j+1})t] \nonumber \\
    & & +A_{3} \cos(\Delta \omega t - \phi_d). 
\end{eqnarray}
$A_0$ accounts for any remaining constant contribution in the coupling.
The $A_1$, $A_2$ components activate local and two-site squeezing terms, respectively. The $A_3$ term activates boson hopping. 
Assuming $\Delta \omega$, $\omega_j$ $\gg$ $A_{j=0,1,2,3}$ we can neglect non-resonant terms in a rotating-wave approximation. We find that $V(t)$ leads to the Hamiltonian part of our array of parametric oscillators with parameters:
$J = A_3/2$,   
$\phi = \phi_d $, 
$g_{s}  =  A_1/2$ and  
$g_{c} = A_2/2$.    

\section{Conclusions\label{sec:Conclusions}}
We have studied the phenomena of topological amplification in arrays of parametric oscillators and have found two qualitatively different driven-dissipative topological phases, one dominated by parametric driving, which requires local dissipation, and other where collective pump dominates.
The existence of an additional phase of amplification with $W_1(\omega)=0$ allowed us to demonstrate that exponential gain due to zero-energy modes does not necessarily imply the presence of topologically protected amplification, which we confirmed by studying the resilience to disorder of the different phases.
In addition, we have studied the stability of the system, finding that the presence of parametric terms makes the stability of the amplifier a size-dependent quantity.
This means that the design of the amplifier must balance several ingredients: array size, strength of parametric terms, dissipation, and gap size.
However, we have shown that it is always possible to find realistic and stable dissipative topological phases with good amplification properties.

Regarding the physical properties of each phase, we have found that the one dominated by parametric driving can be used to generate states with one quadrature squeezed and the other amplified.
In contrast, the topological phase with collective pump amplifies both quadratures.
This difference between the two topological phases indicates that they can be used for different technological applications, with the advantage that they are both present in the same physical setup.
We have found that the gain and noise-to-signal ratio are excellent in both topological phases, and we have been able to show analytically that the dissipative BdG topological phase can reach quantum-limited noise and maximum squeezing near the critical point $\gamma/J\simeq 2$.
Actually, the qualitatively similar behavior of signal-to-noise ratio in both topological phases seems to indicate that the neighborhood of critical points is an interesting working point for topological amplifiers.
There, the amplifier displays great performance, but in practice this must be balanced to avoid saturation~\cite{Ramos2022}.\\
These ideas can be immediately implemented with current superconducting circuit technology~\cite{Ramos2022} or with trapped ions.
From a technological point of view, realizing directional broadband amplification that can be integrated on chip, without bulky and lossy isolators, may help to scale up quantum devices.

As future prospects of our results, it would be interesting to study the capabilities of arrays of parametric oscillators as single photon detectors.
In addition, the use of topological TWPA in the generation of two-mode squeezing is an intriguing direction to pursue~\cite{PhysRevLett.128.153603}.\\
Finally, the application of these ideas to higher-dimensional topological systems might lead to novel effects such as amplifiers with spatial tunability or interesting quadrature properties.
For this, notice that the theory developed in this work can be straightforwardly applied to 2D systems, where the zoo of topological phases now includes chiral or helical edge states.
\begin{acknowledgments}
We acknowledge financial support from the Proyecto Sin\'ergico CAM 2020 Y2020/TCS-6545 (NanoQuCo-CM), the CSIC Interdisciplinary Thematic Platform (PTI+) on Quantum Technologies (PTI-QTEP+) and from Spanish project PGC2018-094792-B-100(MCIU/AEI/FEDER, EU). T.R. further acknowledges support from the Juan de la Cierva fellowship IJC2019-040260-I.
\end{acknowledgments}
\bibliographystyle{apsrev4-2}
\bibliography{main.bib}
\clearpage
\newpage

\onecolumn
\appendix
\section{Input-output description of the TWPA\label{sec:Appendix-InputOutput}}
To describe the TWPA in terms of the input-output formalism~\cite{Zoller-QuantumNoise,PhysRevA.103.033513}, we start by writing the quantum Langevin equation for the photonic operators:
\begin{equation}
    \partial_t a_j = -i\sum_{l=0}^{N-1}\left[(J_{j,l}+i\Gamma_{j,l})a_l + K_{j,l}a_l^\dag \right]+ \xi_j^{\rm in}\label{eq:App-LangevinEq},
\end{equation}
with the coupling terms:
\begin{eqnarray}
    \Gamma_{j,l}&=&\left(\frac{4P-\gamma}{2}\right)\delta_{j,l}+P\left(\delta_{j,l+1}+\delta_{j,l-1}\right)\label{eq:App-coupling1},\\
    J_{j,l}&=&J\left(e^{-i\phi}\delta_{j,l+1}+e^{i\phi}\delta_{j,l-1}\right)\label{eq:App-coupling2},\\
    K_{j,l}&=&g_s+g_c\left(\delta_{j,l+1}+\delta_{j,l-1}\right) .
    \label{eq:App-coupling3}
\end{eqnarray}
The total noise operators in Eq.~\eqref{eq:App-LangevinEq} are (do not confuse the $p_j^{\text{in}}$ operators with the quadrature operators $P_j^{\text{out}}$ in the main text):
\begin{equation}
\xi_j^{\rm in}(t)=-\sqrt{\gamma} a^{\rm in}_j(t)+\sum_{m} \sqrt{\bar{P}_m}R_{mj}^\ast p_m^{\rm in}{}^\dag(t),
\end{equation}
where $R_{ml}$ is a unitary matrix obtained from the eigenvalue decomposition:
\begin{equation}
P_{j,l} = \sum_{m} \bar{P}_m R_{m j}^* R_{ml},\label{eq:App-Simplify-gamma}
\end{equation}
and $\bar{P}_{m}$ are real eigenvalues describing the collective rates for incoherent pumping associated to the noise operators $p_m^{{\rm in} \dag}(t)$.

To close the system of equations produced by Eq.~\eqref{eq:App-LangevinEq}, one just needs to define BdG spinors 
$\vec{a}(t)=[a_0(t),a_1(t),\ldots,a_0^\dag(t),a_1^\dag(t),\ldots]^{T}$ and $\vec{\xi}_{\rm in}(t)=[\xi_0^{\rm in}(t),\xi_1^{\rm in}(t),\ldots,\xi_0^{\rm in}{}^\dag(t),\xi_1^{\rm in}{}^\dag(t),\ldots]^T$, to find:
\begin{align}
     \partial_t\vec{a}(t)= -iH_{\rm nh}\vec{a}(t) +  \vec{\xi}_{\rm in}(t),
\end{align}
where the $2N \times 2N$ non-Hermitian dynamical matrix reads:
\begin{align}
    H_{\rm nh}=\begin{pmatrix}
    J+i\Gamma & K \\
    -K^\ast & -J^\ast+i\Gamma^\ast
    \end{pmatrix}.\label{eq:App-EffectiveH1}
\end{align}
and each block is an $N\times N$ matrix with elements from Eqs.~\eqref{eq:App-coupling1}-\eqref{eq:App-coupling3}.
We can formally solve this system of equations in terms of the Green's function 
\begin{equation}
G(\omega)=(\omega - H_{\rm nh})^{-1},
\label{eq:App-G.w}
\end{equation} 
after a Fourier transform of the operators 
$\tilde{a}_j(\omega) = (2\pi)^{-1/2}\int dt e^{i\omega t} a_j(t)$.
The solution can be written as: 
\begin{align}
    \vec{a}(\omega)=iG(\omega)\vec{\xi}_{\rm in}(\omega),\label{eq:App-FourierSpace}
\end{align}
where $\vec{a}(\omega)=[a_0(\omega),a_1(\omega),\ldots,a_0^\dag(-\omega),a_1^\dag(-\omega),\ldots]^T$ and $\vec{\xi}_{\text{in}}(\omega)=[\xi^{\rm in}_0(\omega),\xi^{\rm in}_1(\omega),\ldots,\xi_0^{\rm in}{}^\dag(-\omega),\xi_1^{\rm in}{}^\dag(-\omega),\ldots]^T$. 
From Eq.~(\ref{eq:App-FourierSpace}), we can write the explicit solution for the Fourier transform of the operators as:
\begin{equation}
    a_j (\omega) = i\sum_{l=0}^{N-1} \left[ G_{j,l}(\omega)\xi_l^{\rm in}(\omega)+G_{j,N+l}(\omega)\xi_l^{\rm in}{}^\dag(-\omega)\right].\label{eq:App-linearsolution}
\end{equation}
Finally, from Eq.~(\ref{eq:App-linearsolution}), the solution for $G(\omega)$ and the input-output relation:
\begin{equation}
    a^{\rm out}_j(\omega) = a^{\rm in}_j(\omega)+\sqrt{\gamma}a_j(\omega)\label{eq:App-input-output},
\end{equation}
one can calculate arbitrary correlation functions of the output field.

To characterize the amplification properties, we are interested in the propagation of an input signal, inserted at the edge, and detected at a particular site, $j$.
For an input signal given by a coherent state with amplitude $\alpha$ and frequency $\omega_d$, we can write the input field as the average value and fluctuations:
\begin{equation}
    a_j^{\text{in}}(\omega)=\langle a_j^{\text{in}}(\omega) \rangle + \delta a_j^{\text{in}}(\omega),
\end{equation}
where $\langle a_j^{\text{in}}(t) \rangle=\alpha \delta_{j,0}e^{-i\omega_d t}$.
Similarly, the output field can be written in terms of its average value and fluctuations:
\begin{equation}
    a_j^{\text{out}}(\omega)=\langle a_j^{\text{out}}(\omega) \rangle + \delta a_j^{\text{out}}(\omega),
\end{equation}
where the average value can be calculated using Eq.~\eqref{eq:App-input-output}, to give:
\begin{align}
    \langle a_{j}^{\text{out}}\left(\omega\right)\rangle=&\alpha\delta\left(\omega-\omega_{d}\right)\left[\delta_{j,0}-i\gamma G_{j,0}\left(\omega\right)\right]\nonumber\\
    &-i\alpha^{*}\delta\left(\omega+\omega_{d}\right)\gamma G_{j,N}\left(\omega\right),
\end{align}
which illustrates the presence of an output signal at two different frequencies, $\pm\omega_d$, typically referred to as signal and idler, respectively.
From this, we can define the gain of the amplifier at signal frequency  and site $j\neq0$ as:
\begin{equation}
    \mathcal{G}_j(\omega)=\gamma^2 |G_{j,0}(\omega)|^2 .
    \label{eq:App-Gain}
\end{equation}
Analogously, as we are also interested in the noise properties, we can calculate the noise added by the amplifier at site $j$,
$n^{\text{amp}}_j (\omega) = \langle \delta a_{j}^{\text{out}\dagger}\left(\omega\right)\delta a_{j}^{\text{out}}\left(\omega\right)\rangle$:
\begin{align}
    n^{\text{amp}}_j (\omega)=&\gamma^2 \sum_{l=0}^{N-1}|G_{j,N+l}\left(\omega\right)|^{2}\nonumber\\
    &+\gamma \sum_{l,l^{\prime}=0}^{N-1}P_{l^{\prime},l}G_{j,l}^{*}\left(\omega\right)G_{j,l^{\prime}}\left(\omega\right),
\end{align}
and the normalized noise-to-signal ratio:
\begin{equation}
    n^\text{add}_j(\omega)=\frac{n_{j}^{\text{amp}}(\omega)}{\mathcal{G}_{j}(\omega)} .
    \label{eq:App-signal-noise-ratio}
\end{equation}
\section{Topological analysis\label{sec:Appendix-Topology}}
We start from the expression for $H_{\text{nh}}$ in terms of Pauli matrices, Eq.~\eqref{eq:H_nh}.
Then, we can write the doubled Hamiltonian in terms of the Pauli matrices acting on each subspace:
\begin{equation}
    \mathcal{H}(k,\omega)=\sum_{j,l}h_{j,l}(k,\omega)\sigma_{j}\otimes\tau_{l},
\end{equation}
where $\sigma_j$ acts on the Nambu spinor subspace and $\tau_j$ on the additional degree of freedom introduced to write $\mathcal{H}$.
If we now consider the presence of TRS, which fulfills: $\mathcal{T}\mathcal{H}(k,\omega)^{*}\mathcal{T}^{-1}=\mathcal{H}(-k,\omega)$, one finds that it is always broken for $\phi\neq 0\mod(\pi)$, because the coefficient $h_{0,x}(k,\omega)=\omega+2J\sin(k)\sin(\phi)$ has a relative sign change for $k\to-k$ which makes impossible to find a transformation that fulfills the criteria for TRS.
This means that PHS also must be broken, in order to have chiral symmetry, which makes the system to belong to the AIII class, which is characterized by a $\mathbb{Z}$ topological invariant in 1D.

On the contrary, if $\phi=0\mod(\pi)$, TRS is given by $\mathcal{T}=\sigma_z\otimes\tau_x$ (with $\mathcal{T}^2=+1$) and PHS results in $\mathcal{C}=\sigma_z\otimes\tau_y$ (with $\mathcal{C}^2=-1$).
In that case the system belongs to the CI class, which is trivial in 1D.
The invariant for the AIII class:
\begin{equation}
    W_{1}(\omega)=\int_{-\pi}^{\pi}\frac{dk}{4\pi i}\textrm{tr}\left[\tau_z\mathcal{H}(k,\omega)^{-1}\partial_k\mathcal{H}(k,\omega)\right],\label{eq:App-Winding}
\end{equation}
correctly predicts the existence of topological amplification, however, it would be interesting to understand the change produced by $P_{j,l}$, which makes trivial edge states to become topologically protected, without changing the topological class.

For that, and due to the chiral symmetry, we can focus on one of the blocks $\omega-H_{\text{nh}}(k)$, where:
\begin{align}
    H_{\text{nh}}\left(k\right) =& \left[ -2J\sin\left(k\right)\sin\left(\phi\right)-i\frac{\gamma}{2}+4i P\cos^{2}\left(\frac{k}{2}\right)\right]\sigma_0\nonumber\\
    &+i\left[ g_s+2 g_c \cos(k) \right]\sigma_{y}\nonumber\\
    &+\left[\Delta+2J\cos(k)\cos(\phi)\right]\sigma_{z},
\end{align}
According to the standard classification of non-hermitian matrices~\cite{PhysRevX.9.041015}, $H_\text{nh}(k)$ has TRS implemented by $\mathcal{T}_{-}=\sigma_x$, which makes it belong to the $\text{D}^{\dagger}$ class in 1D.
Its complex eigenvalues are:
\begin{align}
    E_{\pm}(k)=&-2J\sin\left(k\right)\sin\left(\phi\right)-i\frac{\gamma}{2}+4i P\cos^{2}\left(\frac{k}{2}\right)\\
    &\pm \sqrt{\left[\Delta+2J\cos(k)\cos(\phi)\right]^2-\left[ g_s+2 g_c \cos(k) \right]^2},\nonumber
\end{align}
and each of them results in a point gap, which is known to give a non-vanishing Winding number when enclosing the origin:
\begin{equation}
    W_{\pm}(\omega)=\frac{1}{2\pi i}\int_{-\pi}^{\pi} \partial_k \log\left[\omega-E_{\pm}(k)\right]dk
\end{equation}
Furthermore, the topological invariant from Eq.~\eqref{eq:App-Winding} can be re-written in terms of $H_{\text{nh}}$ as:
\begin{equation}
    W_1(\omega)=\int_{-\pi}^{\pi}\frac{dk}{2i\pi}[\omega-H_{\text{nh}}(k)]^{-1}\partial_{k}[\omega-H_{\text{nh}}(k)],
\end{equation}
and after some manipulations, it can be reduced to:
\begin{equation}
    W_1(\omega)=W_{+}(\omega)+W_{-}(\omega) .
\end{equation}
where we have used the spectral decomposition for $H_{\text{nh}}(k)$ and noticed that only the contribution from the winding of the eigenvalues contributes in this case.
\section{General Green's function from decimation\label{sec:App-Decimation}}
When we rewrite Eq.~\eqref{eq:Propagator-Edge} in terms of the eigenvalues and projectors of the matrix $G_{0,0}(\omega)\mathcal{V}_{-}$:
\begin{equation}
    G_{0,0}(\omega)\mathcal{V}_{-}=\sum_{\alpha=\pm}\lambda_{\alpha}(\omega)P_\alpha(\omega),
\end{equation}
we find the following expression for the Green's function:
\begin{equation}
    G_{j,0}(\omega)=\sum_{\alpha=\pm}e^{\zeta_{\alpha}(\omega)j}P_{\alpha}(\omega) G_{0,0}(\omega) .
\end{equation}
where we have defined the inverse coherence length $\zeta_{\pm}(\omega)=\log[\lambda_{\pm}(\omega)]$ and the projectors fulfill the standard condition $P_{n}P_{m}=P_{n}\delta_{n,m}$.

Notice that for non-hermitian matrices one needs to be careful with the meaning of the eigenvalues. One could adopt the formalism of ref.~\cite{Brody_2013}, based on a biorthogonal basis of eigenvectors, or as it is our case, skip the discussion of the eigenvectors and directly work with projectors.
For that, one just needs to determine the eigenvalues of the non-hermitian matrix, e.g., from its characteristic polynomial $\det\left[G_{0,0}(\omega)\mathcal{V}_{-}-\lambda \mathbf{1}\right]=0$, and then solve the matrix equations:
\begin{align}
    \mathbf{1}&=P_{+}+P_{-}\\
    G_{0,0}(\omega)\mathcal{V}_{-}&=\lambda_{+}P_{+}+\lambda_{-}P_{-}
\end{align}
Its solution immediately gives the matrix form of the projectors, in terms of $\lambda_{\pm}$ and $G_{0,0}(\omega)\mathcal{V}_{-}$:
\begin{equation}
    P_{\pm}=\pm\frac{G_{0,0}(\omega)\mathcal{V}_{-}-\lambda_{\mp}\mathbf{1}}{\lambda_{+}-\lambda_{-}} .
\end{equation}
\section{Analytical expressions in the semi-infinite limit\label{sec:Noise}}
In this Appendix we detail the calculation of the amplifier properties in the semi-infinite limit.

To study the topological phase with local dissipation only, and calculate the different observables in terms of the Green's function, we start from Eq.~\eqref{eq:GreenFunction1} and the solution for the surface Green's function in Eq.~\eqref{eq:non-linearEq}, for the case $g_{s,c}/J=1$, $\phi=\pi/2$ and $\Delta/J=0$:
\begin{equation}
    G_{0,0}\left(\omega\right) = \frac{\left(\begin{array}{cc}
\omega+i\frac{\gamma}{2} & J\\
-J & \omega+i\frac{\gamma}{2}
\end{array}\right)}{J^{2}+\left(\omega+i\frac{\gamma}{2}\right)^{2}} .
\end{equation}
As the relevant quantities for the calculation of an arbitrary Green's function are just $G_{0,0}(\omega)\mathcal{V}_{\pm}$, we can express them in terms of their eigenvalues and projectors.
In particular for our choice of parameters, we find that one of the eigenvalues is always zero, which means that we can write:
\begin{align}
    G_{0,0}(\omega)\mathcal{V}_{-}=&\lambda_{1}P_{1},\\
    G_{0,0}(\omega)\mathcal{V}_{+}=&\lambda_{2}P_{2},
\end{align}
where the eigenvalues and projectors are given by:
\begin{align}
    \lambda_{1}=&\frac{2iJ}{\omega-i\left(J-\frac{\gamma}{2}\right)},P_{1}=\frac{1}{2}\left(\begin{array}{cc}
1 & -i\\
i & 1
\end{array}\right) ,\\
    \lambda_{2}=&\frac{-2iJ}{\omega+i\left(J+\frac{\gamma}{2}\right)},P_{2}=\frac{1}{2}\left(\begin{array}{cc}
1 & i\\
-i & 1
\end{array}\right) .
\end{align}
For the calculation of the gain at site $j$ we are just required to find $G_{j,0}(\omega)$:
\begin{align}
    \mathcal{G}_j(\omega)=&\gamma^{2} |G_{j,0}(\omega)|^2=\gamma^2 |[G_{0,0}(\omega)\mathcal{V}_{-}]^{j}G_{0,0}(\omega)|^2\nonumber\\
    =& \gamma^{2} |e^{j\log (\lambda_{1})} P_{1}G_{0,0}(\omega)|^2\nonumber\\
    =& \frac{\gamma^{2}4^{j-1}J^{2j}}{\left[\omega^{2}+\left(\frac{\gamma}{2}-J\right)^{2}\right]^{j+1}} .
\end{align}

One can proceed in a similar way for the calculation of the noise-to-signal ratio.
The main difference is that now the projector $P_2$ and the eigenvalue $\lambda_{2}$, are also required.
From Eq.~\eqref{eq:signal-noise-ratio} one can see that the sum over $l$ requires to explicitly evaluate Eq.~\eqref{eq:GreenFunction1} for both, $l\leq j$ and $l\geq j$ separately.
Particularizing for each case the expression for the Green's function, we find the following:
\begin{align}
\label{eq:Appendix-GF1}
    G_{j,l\leq j}=& P_{1}G_{0,0}e^{\left(j-l\right)\log\lambda_{1}}\\\nonumber
    &+P_{1}P_{2}G_{0,0}\sum_{r=0}^{l}e^{\left(j-r\right)\log\lambda_{1}+\left(l-r\right)\log\lambda_{2}}, \label{eq:Appendix-GF2}\\
    G_{j,l\geq j}=& P_{2}G_{0,0}e^{\left(l-j\right)\log\lambda_{2}}\\\nonumber
    &+P_{1}P_{2}G_{0,0}\sum_{r=0}^{j}e^{\left(j-r\right)\log\lambda_{1}+\left(l-r\right)\log\lambda_{2}} ,
\end{align}
where the sums over $r$ can be easily evaluated. From the final result we just need to select the anomalous component of the Green's function, $G_{j,l}^{1,2}$, which is the required term $G_{j,N+l}(\omega)$.
Finally, we are left with calculating $\sum_{l=0}^{\infty}|G_{j,N+l}(\omega)|^2$, which we separate into $\sum_{l=0}^{j-1}|G_{j,N+l}(\omega)|^2$ and $\sum_{l=j}^{\infty}|G_{j,N+l}(\omega)|^2$ for the corresponding cases in Eqs.~\eqref{eq:Appendix-GF1} and \eqref{eq:Appendix-GF2}.
This results in the analytical value plotted in Fig.~\ref{fig:Gain-Z_2}.
%%%%%%%%%%%%%%%
\end{document}